\documentclass{cta-author}

\newtheorem{theorem}{Theorem}{}
{}
\newtheorem{remark}{Remark}{}
\newtheorem{assumption}{Assumption}{}
\newtheorem{definition}{Definition}{}

\newcommand{\norm}[1]{\vert{#1}\vert}		
\newcommand{\supofseq}[1]{\left\lVert\boldsymbol{#1}\right\rVert}	
\newcommand{\Tr}[1]{Tr\left({#1}\right)}	

\usepackage{mathtools,mathrsfs,amsmath}
\usepackage{booktabs, threeparttable}
\usepackage{caption}
\usepackage{tabularx}

\begin{document}

\title{Robust stability of moving horizon estimation for nonlinear systems with bounded disturbances using adaptive arrival cost}

\author{\au{Nestor N. Deniz$^{1}$}
\au{, Marina H. Murillo$^1$}
\au{, Guido Sanchez$^1$}
\au{, Lucas M. Genzelis$^1$}
\au{, Leonardo L. Giovanini$^1$}
}

\address{\add{1}{Research institute for signals, systems and computational intelligence, Ciudad Universitaria UNL, Ruta Nac. No 168, km 472.4, FICH, 4to Piso
(3000) Santa Fe - Argentina}
\email{ndeniz@sinc.unl.edu.ar}}

\begin{abstract}
In this paper, the robust stability and convergence to the true state of moving horizon estimator based on an adaptive arrival cost are established for nonlinear detectable systems. Robust global asymptotic stability is shown for the case of non-vanishing bounded disturbances whereas the convergence to the true state is proved for the case of vanishing disturbances. Several simulations were made in order to show the estimator behaviour under different operational conditions  and to compare it with the state of the art estimation methods.
\end{abstract}

\maketitle

\section{Introduction}
State estimation plays a fundamental role in feedback control, system monitoring and system optimization because noisy measurements is the only information available from the system. Several methods have been developed for accomplishing such task (see \cite{jazwinski2007stochastic};  \cite{crassidis2004optimal}; among others). All these methods have been developed upon the assumption on the knowledge of noises and model of the system, as well as, the absence of constraints.

In practice, these assumptions are not easily satisfied and research efforts were focused on approaches that do not relay on such requirements (see \cite{li1997linear}, \cite{sayed2001framework}, \cite{blanchini2008set}, among others). For example, an $H_{\infty}$ filter is designed minimizing the $H_{\infty}$ norm of the mapping between disturbances and estimation error. In \cite{li1997linear}, \cite{el1997robust} and \cite{hu2009improved} an approach that solves a least-square estimation problem is introduced. Both methods are based on the adequate selection of the uncertainty model instead of relying on statistical assumptions on noises. In these approaches, uncertainty models are formulated based on the available information of the system. In the same way, robust estimation algorithms based on as min-max robust filtering, set-valued estimation and guaranteed cost paradigm, have attracted the attention of the research community (see \cite{sayed2001framework}, \cite{zhu2002design}).

Building on the success on moving horizon control, moving horizon estimation (MHE) has attracted attention of researchers since the pioneering work of \cite{jazwinski1968limited} (see also \cite{schweppe1973uncertain}, \cite{rao2001constrained} and \cite{rao2003constrained}). The interest in such estimation methods stems from the possibility of dealing with limited amount of data, instead of using all the information available from the beginning, and the ability to incorporate constraints. In recent years, both theoretical properties of various MHE schemes as well as efficient computational methods for real-time implementation have been studied (see \cite{alessandri2005robust}, \cite{alessandri2008moving}, \cite{alessandri2012min}, \cite{garcia2016new}, \cite{sartipizadeh2016computationally}, \cite{sanchez2017adaptive}). In particular, it is of interest to establish robust stability and estimate convergence properties. In recent years several results have been obtained for different algorithms, advancing from idealistic assumptions (observability and no disturbances) to realistic situations (detectability and bounded disturbances). 

For nonlinear observable systems, \cite{rao2003constrained} established the asymptotic stability of the estimation error for the standard cost function.  Furthermore, if the disturbances are asymptotically vanishing the estimation error is robust asymptotically stable and it asymptotically converges to zero (\cite{rawlings2009model}-\cite{rawlings2012optimization}). \cite{alessandri2008moving} and \cite{alessandri2010advances} proposed an estimation scheme, based on least-square cost function of the estimation residuals, that guaranteed the boundedness of estimation error for observable systems subject to bounded additive disturbances. Finally, for the general case of nonlinear detectable systems subject to bounded disturbances, \cite{ji2016robust} and \cite{muller2017nonlinear} showed the robust global asymptotic stability (RGAS) and convergence of estimation error in case of bounded or vanishing disturbances, respectively. In these works, the least-square objective function was modified by adding a max-term. \cite{ji2016robust} established RGAS for the full information estimator while \cite{muller2017nonlinear} established RGAS and convergence for the moving horizon estimator. Furthermore, for a particular choice of the weights of the objective function, \cite{muller2017nonlinear} established these results for the least-squares type objective function.

This paper introduces the RGAS and convergence analysis for the moving horizon estimator based on adaptive arrival cost proposed 
in \cite{sanchez2017adaptive} in the practical case of nonlinear detectable systems subject to bounded disturbances. To establish robust stability properties for MHE it is crucial that the prior weighting in the cost function is chosen properly. In various schemes the necessary assumptions in the prior weighting are difficult to verify (\cite{rao2003constrained}, \cite{rawlings2009model}), while in others can be verified a prior \cite{muller2017nonlinear}. In the MHE scheme analysed in this work, the assumption on the prior weighting can be verified a prior by design. Furthermore, the disturbances gains become uniform (i.e., they are valid independent of $N$), allowing to extend the stability analysis to full information estimators with least-square type cost functions. 

The rest of the paper is organized as follows: Section 2 introduces the notation, definitions and properties that will be used through the paper. Section 3 presents the main result and shows its connections with previous stability analysis. Section 4 discusses simple examples, previously used in the literature, with the purpose of illustrating the concepts and also in order to show the difference with others MHE algorithms. Finally, Section 5 presents conclusions.



\section{Preliminaries and setup}

\subsection{Notation}
Let $\mathbb{Z}_{\left[a,b \right]}$ denotes the set of integers in the interval $\left[ a,b \right] \subseteq \mathbb{R} \textnormal{, and } \mathbb{Z}_{\geq a}$ denotes the set of integers greater or equal to $a$. Boldface symbols denote sequences of finite or infinite length, i.e., $\boldsymbol{w} \coloneqq \{ w_{k_1}, \ldots, w_{k_2}  \} \textnormal{ for some } k_1, k_2 \in \mathbb{Z}_{\geq 0} \textnormal{ and } k_1 < k_2$, respectively. We denote $x_{j\vert k}$ as the finite sequence $\boldsymbol{x}$ given at time $k \in \mathbb{Z}_{\geq 0} \textnormal{ and } j \in \left[k_1, k_2 \right]$. 
By $\norm{x}$ we denote the Euclidean norm of a vector $x \in \mathbb{R}^n$. Let $\supofseq{x} \coloneqq \sup_{k\in \mathbb{Z}_{\geq 0}} \norm{x_k} $ denote the supreme norm of the sequence $\boldsymbol{x} \textnormal{ and } \supofseq{x}_{\left[a, b \right]} \coloneqq \sup_{k\in \mathbb{Z}_{\left[a, b \right]}} \norm{x_k} $ . A function $\gamma : \mathbb{R}_{\geq 0} \rightarrow \mathbb{R}_{\geq 0}$ is of class $\mathcal{K}$ if $\gamma$ is continuous, strictly increasing and $\gamma \left(0\right) = 0$ . If $\gamma$ is also unbounded, it is of class $\mathcal{K}_\infty$. A function $\zeta : \mathbb{R}_{\geq 0} \rightarrow \mathbb{R}_{\geq 0}$ is of class $\mathcal{L}$ if $\zeta \left( k \right)$ is non increasing and $\lim_{k \rightarrow \infty} \zeta\left(k \right) = 0$. A function $\beta : \mathbb{R}_{\geq 0}\times\mathbb{Z}_{\geq 0} \rightarrow \mathbb{R}_{\geq 0}$ is of class $\mathcal{KL}$ if $\beta \left(\cdot,k \right)$ is of class $\mathcal{K}$ for each fixed $k \in \mathbb{Z}_{\geq 0}$, and $\beta \left(r,\cdot \right)$ of class $\mathcal{L}$ for each fixed $r \in \mathbb{R}_{\geq 0}$. 

The following inequalities hold for all $\beta \in \mathcal{KL}, \; \gamma \in \mathcal{K} \textnormal{ and } a_j \in \mathbb R_{\geq 0} \textnormal{ with } j \in \mathbb{Z}_{\left[1,n \right]}$
\begin{equation}		\label{property_1}
  \begin{split}
    \gamma\left(a_1 + a_2 + \ldots + a_n \right) &\leq \gamma\left (n a_1\right) + \ldots + \gamma\left (n a_n\right), \\
    \beta\left(a_1 + a_2 + \ldots + a_n, k \right) &\leq \beta\left(n 
    a_1, k\right) + \ldots + \beta\left(n 
    a_n, k\right).
  \end{split}
\end{equation}
The preceding inequalities hold since $\max\{a_j \}$ is included in the sequence $\{ a_1, a_2, \ldots, a_n \}$ and $\mathcal{K}$ functions are non-negative strictly increasing functions. 

\textbf{Bounded sequences:} A sequence $\boldsymbol{w}$ is bounded if $\supofseq{w}$ is finite. The set of bounded sequences $\boldsymbol{w}$ is denoted as $\mathcal{W}\left(w_{\max}\right) \coloneqq \{w : \supofseq{w} \leq w_{\max} \}$ for some $w_{\max} \in \mathbb{R}_{\geq 0}$

\textbf{Convergent sequences:} A bounded infinite sequence $\boldsymbol{w}$ is convergent if $\norm{w_k} \rightarrow 0$ as $k \rightarrow \infty$. Let denote the set of convergent sequences $\mathcal{C}$:
\begin{equation*}
  \begin{array}{rcl}
    \mathcal{C}_w \coloneqq \{ \boldsymbol{w} \in \mathcal{W}\left(w_{\max}\right) 
    \vert \, \boldsymbol{w}\text{ is convergent} \}
\end{array}
\end{equation*}
Analogously, $\mathcal{C}_v$ is defined for the sequence \textbf{$\boldsymbol{v}$}.

\subsection{Problem statement} %
Let us consider the state estimation problem for nonlinear discrete time systems of the form
\begin{equation}		\label{eq_nonlinsys}
  \begin{split}
	x_{k+1} &= f\left(x_k, w_k\right), \qquad x_0 = \mathtt{x}_0\\
	y_{k} 	&= h\left(x_k\right) + v_k,
  \end{split}
\end{equation}

where $x_{k} \in \mathcal{X} \subseteq \mathbb{R}^n, w_{k} \in \mathcal{W} \subseteq \mathbb{R}^p, y_k \in \mathcal{Y} \subseteq \mathbb{R}^m, v_{k} \in \mathcal{V} \subseteq \mathbb{R}^m$  are the state, process noise, measurement and estimation residuals vectors, respectively. The process disturbance $w_k$ and estimation residuals $v_k$ are unknown but assumed to be bounded, i.e, $\boldsymbol{w} \in \mathcal{W}\left( w_{\max} \right) \textnormal{ and } \boldsymbol{v} \in \mathcal{V}\left( v_{\max} \right)$ for some $w_{\max}, v_{\max} \in \mathbb{R}_{\geq 0}$. $\mathcal{X}, \mathcal{Y}, \mathcal{W} \textnormal{ and } \mathcal{V}$ are compact and convex sets with the null vector $\mathbf{0}$ belongs to them. In the following we assume that $f : \mathbb{R}^n \times \mathbb{R}^p \rightarrow \mathbb{R}^n$ is continuous, locally Lipschitz on $x_k$ and $h : \mathbb{R}^n \rightarrow \mathbb{R}^m$ is continuous.
The solution to the system \eqref{eq_nonlinsys} at time $k$ is denoted by $x\left(k; \mathtt{x}_0, \boldsymbol{w} \right)$, with initial condition $\mathtt{x}_0$ and process disturbance sequence $\boldsymbol{w}$. Furthermore, the initial condition $\mathtt{x}_0$ is unknown, but a prior knowledge $\bar{x}_0$ is assumed to be available and its error is assumed to be bounded, i.e., $\bar{x}_0 \in \mathcal{X}_0 \coloneqq \{ \bar{x}_0 : \norm{\mathtt{x}_0 - \bar{x}_0} \leq e_{\max} \}$, $\mathcal{X}_0 \subseteq \mathcal{X}$. 

The solution of the estimation problem aims to find at time $k$ an estimate  $\hat{x}_{k  \vert k}$ of the current state $x_k$ minimizing a performance metric using by the MHE. At each sampling time $k$, given the previous $N$ measurements $\boldsymbol{y} := \{ y_{k-N}, \ldots, y_{k-1} \}$, the following optimization problem is solved
\begin{equation}  \label{eq:mhe1}
  \begin{array}{c}
    \underset{\hat{x}_{k-N|k}, \boldsymbol{\hat{w}}_{j\vert k} }
    {\operatorname{min}} \, \Psi \coloneqq \Gamma_{k-N \vert k}
    \left(\hat{x}_{k-N\vert k}   \right) + \sum\limits_{j=k-N}^{k-1}
    \ell \left(  \hat{w}_{j\vert k}, \hat{v}_{j\vert k}\right) \\
    \text{s.t.} \left\{
    \begin{array}{l}
      \begin{array}{rll} 
\hat{x}_{j+1|k} 	 =& f\left(\hat{x}_{j\vert k}, \hat{w}_{j\vert k}\right) ,& j \in \mathbb{Z}_{\left[k-N, k-1\right]}	\\
y_{j} 		 =& h\left(\hat{x}_{j\vert k}\right) + \hat{v}_{j\vert k} ,& 	\\
      \end{array} \\
\hat{x}_{j|k} \in \mathcal{X}, \, \hat{w}_{j|k} \in \mathcal{W}, \ \hat{v}_{j|k} \in \mathcal{V},
    \end{array} \right.
  \end{array}
\end{equation}
where $\hat{x}_{k-j\vert k}$ is the optimal estimated and $\hat{w}_{j\vert k}$ is the optimal process noise estimate at sample $k-j \quad j=0, 1, \ldots, N$ based on measurements $y_{k-j}$ available at time $k$. The process noise $\boldsymbol{\hat{w}}_{j\vert k} \coloneqq \{\hat{w}_{k-N\vert k}, \ldots, \hat{w}_{k-1\vert k} \}$ and $\hat{x}_{k-N\vert k}$ are the optimization variables. The stage cost $\ell \left(w_{j\vert k}, v_{j\vert k} \right)$ penalizes the estimated process noise sequence $\boldsymbol{\hat{w}}_{j\vert k}$ and the estimation residuals $\boldsymbol{\hat{v}}_{j\vert k} = \boldsymbol{y}_{j} - h\left( \boldsymbol{\hat{x}}_{j\vert k}\right)$, while $\Gamma_{k-N} \left(\hat{x}_{k-N\vert k} \right)$ penalizes the prior estimated $\hat{x}_{k-N\vert k}$. The adequate choice of $\ell\left( \cdot \right)$ and $\Gamma_{k-N} \left( \cdot \right)$, and their parameters, allows to ensure the robust stability of the estimator \cite{muller2017nonlinear}. While the estimation window is not full, $k \leq N$, problem \eqref{eq:mhe1} can be reformulated and solved as a full information problem
\begin{equation*}                   \label{eq:mhe2}
\begin{array}{c}
\underset{\hat{x}_{k-N|k}, \boldsymbol{\hat{w}} }
{\operatorname{min}} \, \Psi \coloneqq \Gamma_{0\vert k}\left(\hat{x}_{0\vert k} \right) + \sum\limits_{j=0}^{k-1} \ell\left( \hat{w}_{j\vert k}, \hat{v}_{j\vert k}\right) \\[0.5cm]

\text{s.t.} \left\{
\begin{array}{l}
\begin{array}{rll} 
\hat{x}_{j+1|k} 	 =& f\left(\hat{x}_{j\vert k}, \hat{w}_{j\vert k}\right) ,& j \in \mathbb{Z}_{\left[0, k-1\right]}	\\
y_{j} 		 =& h\left(\hat{x}_{j\vert k}\right) + \hat{v}_{j\vert k} ,&  \\

\end{array} \\
\hat{x}_{j|k} \in \mathcal{X}, \ \hat{w}_{j|k} \in \mathcal{W}, \ \hat{v}_{j|k} \in \mathcal{V},
\end{array} \right.
\end{array}
\end{equation*}
as $k$ increases this problem becomes \eqref{eq:mhe1} for all $k \geq N$.

In previous works, the robust stability of MHE has been achieved by modifying the standard least-square cost function through the inclusion of a $max$--term (\cite{ji2016robust}; \cite{muller2017nonlinear}) or by a suitable choice of the cost's function parameters (\cite{muller2017nonlinear}). Another mechanism to solve this problem is combining a suitable choice of the stage cost $\ell\left(  \hat{w}_{j\vert k}, \hat{v}_{j\vert k} \right)$ with a time--varying prior weight of the form
\begin{equation}
  \Gamma_{k-N \vert k}\left( \hat{x}_{k-N\vert k} \right) = \lVert \, \hat{x}_{k-N\vert k} - \bar{x}_{k-N} \, \rVert_{P^{-1}_{k-N\vert k}},
\end{equation}
whose parameters $\left( P^{-1}_{k-N\vert k},\bar{x}_{k-N} \right)$ are recursively updated using the information available at time $k$ (\cite{sanchez2017adaptive}, 2017). The prior weighting is defined in this way to avoid the introduction of artificial cycling in the estimation process (see \cite{rawlings2009model}). In this approach, the prior weight matrix $P_{k-N\vert k}$ is given by
\begin{equation}  \label{eq:updatePk}
  \begin{split}
  \epsilon_{k-N} & = y_{k-N}-\hat{y}_{k-N \vert k}, 	\\
             N_k & = \left[ 1 + \hat{x}_{k-N\vert k-1}^T\, P_{k-
             N-1} \hat{x}_{k-N\vert k-1} \right]\frac{\sigma}
             {\norm{\epsilon_{k-N}}_2^2} \\
      \alpha_k & = 1 - \frac{1}{N_k},\\
           W_k & = \left[ I - \frac{P_{k-N-1} \hat{x}_{k-
           N\vert k-1} \hat{x}_{k-N\vert k-1}^T}{1+\hat{x}_{k-
           N\vert k-1}^T P_{k-N-1} \hat{x}_{k-N\vert k-1}}
           \right] P_{k-N-1}, \\
           P_{k-N} & = \left\{ \begin{array}{ccc}
           \frac{1}{\alpha_k} W_k & & \text{if } \frac{1}{\alpha_k}
           \Tr{W_k} \leq c, \\
           W_k & & \text{otherwise},
     \end{array} \right.
  \end{split}
\end{equation}
where $\sigma, \, \sigma_w, \, c, \, \lambda \in R_{>0}, \, c>\lambda, \, P_{0}=\lambda I_{n\times n}$ and $\sigma \gg \sigma_w$, where $\sigma_w$ denotes the process noise variance. The prior knowledge of the window $\bar{x}_{k-N}$ is updated using a smoothed estimate (\cite{findeisen1997moving})
\begin{equation}
  \bar{x}_{k-N} = \hat{x}_{k-N\vert k-1}.
\end{equation}
The optimization problem (\ref{eq:mhe1}) can be reformulated  in terms of the initial condition $\hat{x}_0$ and the estimated process noises and the residuals along the entire trajectory as follows
\begin{equation*}
\begin{array}{c}
\underset{\hat{x}_{0|k}, \boldsymbol{\hat{w}} }
{\operatorname{min}} \, \Psi \coloneqq  \sum\limits_{j=k-N}^{k-1} \ell\left( \hat{w}_{j\vert k}, \hat{v}_{j\vert k}\right) + \\
\sum\limits_{j=1}^{k-N-1} \alpha_{k}^{k-N-j} \ell\left( \hat{w}_{j\vert k}, \hat{v}_{j\vert k}\right)
+ \alpha_{k}^{k-N} \Gamma_{0\vert k} \left(\hat{x}_{0|k} \right)
\\[0.5cm]

\text{s.t.} \left\{
\begin{array}{l}
\begin{array}{rll} 
\hat{x}_{j+1|k} 	 =& f\left(\hat{x}_{j\vert k}, \hat{w}_{j\vert k}\right) ,& j \in \mathbb{Z}_{\left[0, k-1\right]}, \alpha \in (0,1]	\\
y_{j} 		 =& h\left(\hat{x}_{j\vert k}\right) + \hat{v}_{j\vert k} ,&  \\
\end{array} \\
\hat{x}_{j|k} \in \mathcal{X}, \ \hat{w}_{j|k} \in \mathcal{W}, \ \hat{v}_{j|k} \in \mathcal{V},
\end{array} \right.
\end{array}
\end{equation*}
This formulation of problem (\ref{eq:mhe1}) allows to explicitly see the effect of past data on the current state estimate $\hat{x}_{k\vert k}$. In this formulation it is easy to see the exponential averaging of these data. Allowing $\alpha$ change in time, the past data has different affects on the current estimates depending on $\hat{x}_{k\vert k}$.

Before proceeding to the development of the main results, we state the main properties and assumptions about the prior weighting $\Gamma_{k-N}$.

The updating mechanism (\ref{eq:updatePk}) is a time-varying filter whose inputs are $\hat{x}_{k-N \vert k-1}\hat{x}_{k-N \vert k-1}^{T}$ and the initial condition $P_0$. It generates  recursively a real-time estimation of $P_{k-N \vert k}$ by updating $P_{k-N-1 \vert k-1}$ with an exponential time-averaging of $\hat{x}_{k-N \vert k-1}\hat{x}_{k-N \vert k-1}^{T}$. The updating mechanism (\ref{eq:updatePk}) only use data and it does not rely on a model of the system. The sequence $P_{k \vert k } \quad k \geq 0$ is positive definite, it is decreasing in norm and it is bounded. The proof of these properties follows similar steps as in \cite{sanchez2017adaptive}. 

\begin{assumption}		\label{prior weighting assumption}
The prior weighting $\Gamma_{k-N}$ is a continuous function $\Gamma_{k-N}: \mathbb{R}^n \rightarrow \mathbb{R}$ lower bounded by $\underline{\gamma}_p \in \mathcal{K}_\infty{}$ and upper bounded by $\bar{\gamma}_p \in \mathcal{K}_{\infty}$ such that:

\begin{equation}		\label{prior weighting bounds}
\begin{split}
  \underline{\gamma}_p\left(\norm{\hat{x}_{k-N\vert k} - \bar{x}_{k-
  N}} \right) &\leq \Gamma_{k-N}\left(\hat{x}_{k-N\vert k} \right) \\
  \Gamma_{k-N}\left(\hat{x}_{k-N\vert k} \right) &\leq \bar{\gamma}_p \left(\norm{\hat{x}_{k-N\vert k} - \bar{x}_{k-
  N}} \right)
  \end{split}
\end{equation}

for all $\hat{x} \in \mathcal{X}$ and
\begin{equation}
  \underline{\gamma}_p\left(r \right) \geq \underline{c}_p \, r^{a},
  \hspace{0.5cm} \bar{\gamma}_p\left(r \right) \leq \bar{c}_p \, r^{a}.
\end{equation}
where $0 \leq \underline{c}_p \leq \bar{c}_p$ and $a \in R_{\geq 1}$.
\end{assumption}

\noindent Given prior weighting updating scheme \eqref{eq:updatePk} inequality \eqref{prior weighting bounds} satisfies \citep{sanchez2017adaptive}
\begin{equation}		\label{gamma p inequality}
  \norm{P_0^{-1}} r^{a} \leq \Gamma_{k-N}\left( \hat{x}_{k-N\vert k} \right) \leq \norm{P_{\infty}^{-1}} r^{a}.
\end{equation}

\begin{definition}		\label{ioss_definition}
The system \eqref{eq_nonlinsys} is \textit{incrementally input/output-to-state stable} if there exist functions $\beta \in \mathcal{KL}$ and $\gamma_1, \gamma_2 \in \mathcal{K}$ such that for every two initial states $z_1$, $z_2 \in \mathbb{R}^n$, and any two disturbances sequences $\boldsymbol{w_1}, \boldsymbol{w_2}$ the following holds for all $k \in \mathbb{Z}_{\geq 0}$:
\begin{equation}		\label{i_IOSS_propertyerty}
  \begin{split}
    \norm{ x ( k,z_1,\boldsymbol{w_1} ) - x( k,z_2,\boldsymbol{w_2})} &\leq \max \lbrace \beta\left( \norm{z_1 - z_2},k \right), \\
    &\gamma_1\left( \supofseq{w_1 - w_2} \right), \\
    & \quad \gamma_2\left( \supofseq{h\left( x_1 \right) - h\left( x_2 \right)}  \right) \rbrace \\ 
            &\leq \beta ( \norm{z_1 - z_2},k ) + \\
            & \gamma_1 \left( \supofseq{w_1 - w_2} \right) + \\
            & \quad \gamma_2\left( \supofseq{v_1 - v_2} \right)
  \end{split}
\end{equation}
\end{definition}

\noindent This definition combines the concepts of output-to-state-stability (OSS) and input-to-state-stability (ISS). As stated in \cite{sontag1997output}, the notion of IOSS represents a natural combination of the ideas of \textit{strong} observability and ISS, and it was called \textit{detectability} in \cite{sontag1989some} and \textit{strong unboundedness observability} in \cite{jiang1994small}. In addition, the existence of an observer for the system \eqref{eq_nonlinsys}, which is \textit{incrementally input-output-to-state stable} (\textit{i}-IOSS) instead of IOSS (see Remark 24 in \cite{sontag1997output}), is assumed. Note that $\supofseq{h\left( x_1 \right) - h\left( x_2 \right)} = \supofseq{v_1 - v_2}$, since $y_k = h\left( x_k \right) + v_k$ These assumptions will help us to bound the functions involved in the definition of \textit{i}-IOSS and to relate them with the terms of the MHE cost function (stage cost and prior weight).

In the following sections the updating mechanism \eqref{eq:updatePk} and the assumption of \textit{i}-IOSS \cite{sontag2008input} will be used to prove  robust stability of the proposed MHE in the presence of bounded disturbances and convergence to the true state in the case of convergent disturbances. Some assumptions about functions related to system \eqref{eq_nonlinsys} and Definition \ref{ioss_definition}  will be helpful in the sequel.

\begin{assumption}		\label{assumption beta function ineq}
The function $\beta(r,s) \in \mathcal{KL}$ and satisfies the following inequality
\begin{equation}		\label{beta_function_ineq}
  \beta(r,s) \leq c_{\beta}r^p s^{-q}
\end{equation}
for some $c_{\beta} \in \mathbb{R}_{\geq 0}$, $p \in \mathbb{R}_{\geq 0}$ and $q \in \mathbb{R}_{\geq 0}$ and $q\geq p$.
\end{assumption}

\begin{assumption}		\label{stage cost assumption}
The stage cost $\ell\left(\cdot, \cdot \right) : \mathbb{R}^p \times \mathbb{R}^m \rightarrow \mathbb{R}$ is a continuous function bounded by $\underline{\gamma}_w, \underline{\gamma}_v, \bar{\gamma}_w, \bar{\gamma}_v$ $\in \mathcal{K}_\infty{}$ such that the following inequalities are satisfied $\forall w \in \mathcal{W} \textnormal{ and } v \in \mathcal{V}$
  \begin{equation}		\label{stage cost inequalities}
    \underline{\gamma}_w\left(w \right) + \underline{\gamma}_v\left(v
    \right) \leq \ell\left(w, v \right) \leq \bar{\gamma}_w\left(w
    \right) + \bar{\gamma}_v\left(v \right).
  \end{equation}
\end{assumption}

\noindent Functions $\gamma_1$ and $\gamma_2$ from Definition \ref{ioss_definition} are related with the bounds of stage cost $\bar{\gamma}_w, \underline{\gamma}_w, \bar{\gamma}_v$  and $\underline{\gamma}_v$ through the following inequalities
\begin{equation}		\label{gamma1 and gamma2 inequalities}
  \gamma_1\left( 3\underline{\gamma}_w^{-1} \left( r \right)\right) \leq c_1
  r^{\alpha_1}, \;
  \gamma_2\left( 3\underline{\gamma}_v^{-1} \left( r \right)\right) \leq c_2
  r^{\alpha_2}
\end{equation}
for $c_{1}, c_{2}, \alpha_{1},  \alpha_{2} > 0.$ Inequalities \eqref{beta_function_ineq} to \eqref{gamma1 and gamma2 inequalities} were used in previous works (\cite{ji2016robust}; \cite{muller2017nonlinear}). 

In this work, we claim that the proposed estimator holds the property of being robust global asymptotic stable, which is defined as follows.
\begin{definition}
Consider the system described \eqref{eq_nonlinsys} subject to disturbances $\boldsymbol{w} \in \mathcal{W}\left(w_{\max} \right)$ and $\boldsymbol{v} \in \mathcal{V}\left(v_{\max} \right)$ for $w_{\max} \in \mathbb{R}_{\geq 0} $, $v_{max} \in \mathbb{R}_{\geq 0}$ with prior estimate $\bar{x}_0 \in \mathcal{X}\left(e_{\max} \right)$ for $e_{\max} \in \mathbb{R}_{\geq 0}$. The moving horizon state estimator given by equation \eqref{eq:mhe1} with adaptive prior weight is robustly globally asymptotically stable (RGAS) if there exists functions $\Phi \in \mathcal{KL} $ and $\pi_w$, $\pi_v \in \mathcal{K}$ such that for all ${x}_0 \in \mathcal{X}$, all $\bar{x}_0 \in \mathcal{X}_0$, the following is satisfied for all $k \in \mathbb{Z}_{\geq 0}$
\begin{equation}		\label{RGAS def property}
\begin{split}
  \norm{x_k - \hat{x}_k} &\leq \Phi\left(\norm{x_0 - \bar{x}_0},k
  \right) + \pi_w\left(\supofseq{w}_{[0,k-1]} \right) + \\
  & \qquad \pi_v\left(\supofseq{v}_{[0,k-1]} \right).
  \end{split}
\end{equation}
\end{definition}

\noindent We want to show that if system \eqref{eq_nonlinsys} is i-IOSS, then Assumptions \ref{prior weighting assumption}, \ref{assumption beta function ineq} and \ref{stage cost assumption} are fulfilled and the proposed MHE estimator with adaptive arrival cost weight matrix is RGAS. Furthermore, if the process disturbance and measurement noise sequences are convergent (i.e., $\boldsymbol{w}\in \mathcal{C}_w, \boldsymbol{v} \in \mathcal{C}_v $), then $\hat{x}_{k \vert k} \rightarrow x_k$ as $k \rightarrow \infty$.

\section{Robust stability of moving horizon estimation under bounded disturbances}
We are ready to derive the main result: RGAS of the proposed moving horizon estimator with a large enough estimation horizon $\mathcal{N}$ for nonlinear detectable systems under bounded disturbances. Furthermore,  a $\mathcal{K}\mathcal{L}$ function exist such that \eqref{RGAS def property} is valid with this $\Phi, \pi_w$ and $\pi_v$ for all estimation horizon $N \geq \mathcal{N}$.

\begin{theorem}		\label{Theorem_1}
Consider an i-IOSS system \eqref{eq_nonlinsys} with disturbances $\boldsymbol{w} \in \mathcal{W}\left(w_{\max} \right)$, $\boldsymbol{v} \in \mathcal{V}\left(v_{\max} \right)$.  Assume that the arrival cost weight matrix of the MHE problem $\Gamma_{k-N}$ is updated using the adaptive algorithm \eqref{eq:updatePk}. Moreover, Assumptions \ref{prior weighting assumption}, \ref{assumption beta function ineq} and \ref{stage cost assumption} are fulfilled and initial condition $x_0$ is unknown, but a prior estimate $\bar{x}_0 \in \mathcal{X}_0$ is available. Then, the MHE estimator \eqref{eq:mhe1} is $RGAS$. 
\end{theorem}

\textbf{Proof.}
The optimal cost of problem \eqref{eq:mhe1} is given by
\begin{equation*}
\begin{split}
  \Psi_{N}^{*} &= \Psi\left(\hat{x}_{k-N\vert k}^*, \boldsymbol{\hat{w}^*}
  {\left[k-N, k-1 \right]} \right) \\
  &= \Gamma_{k-N} \left(\hat{x}_{k-N \vert k}^* \right) + \sum\limits_{j=k-N}^{k-1} \ell \left( \hat{w}_{j\vert k}^*, \hat{v}_{j\vert k}^*\right),
  \end{split}
\end{equation*}
which is bounded (Assumptions \ref{prior weighting assumption} and \ref{stage cost assumption}) $\forall \, \norm{\hat{w}_{j\vert k}}$ and $\forall \, \norm{\hat{v}_{j\vert k}}$ for all $j  \in \mathbb{Z}_{\left[k-N, k-1 \right]}$ by
\begin{align*}
    \Psi_{N}^{*} & \leq \overline{\gamma}_p\left(\norm{\hat{x}_{k-N\vert k}^* - \bar{x}_{k-N}} \right) + N \overline{\gamma}_w
\left(\norm{\hat{w}_{j\vert k}^*} \right) + N \overline{\gamma}_v \left(\norm{\hat{v}_{j\vert k}^*} \right),  \\
    \Psi_{N}^{*} & \geq \underline{\gamma}_p\left(\norm{\hat{x}_{k-N\vert k}^* - \bar{x}_{k-N}} \right) + N \underline{\gamma}_w
\left(\norm{\hat{w}_{j\vert k}^*} \right) + N \underline{\gamma}_v \left(\norm{\hat{v}_{j\vert k}^*} \right).
\end{align*}
Due optimality, the following inequalities hold $\forall k \in [k-N,k-1]$
\begin{equation}		\label{upper bound for cost function}
  \begin{split}
  \Psi\left(\hat{x}_{k-N\vert k}^*, \boldsymbol{\hat{w}^*} \right) & \leq \Psi\left({x}_{k-N}, \boldsymbol{{w}} \right), \\
  &\leq \bar{\gamma}_p\left(\norm{x_{k-N} - \bar{x}_{k-N}} 
  \right) + \\
  & \qquad N \bar{\gamma}_w\left(\supofseq{w} \right) + N \bar{\gamma}_v\left(\supofseq{v} \right),
 \end{split}
\end{equation}
then, taking into account the lower and upper bounds we have
\begin{equation*}
\begin{split}
\norm{\hat{x}_{k-N\vert k} - \bar{x}_{k-N}} \leq \underline{\gamma}_{p}^{-1}\left(\bar{\gamma}_p\left( \norm{x_{k-N} - \bar{x}_{k-N} }\right) \right. + \\
\left. N \bar{\gamma}_w\left(\supofseq{w} \right) + N \bar{\gamma}_v\left(\supofseq{v} \right) \right).
\end{split}
\end{equation*}
By mean of Assumptions \ref{prior weighting assumption} and \ref{stage cost assumption} the last inequality can be written as follows
\begin{equation*}
  \begin{split}
\norm{\hat{x}_{k-N\vert k} - \bar{x}_{k-N}} &\leq \underline{\gamma}_p^{-1}\left(3\; \bar{\gamma}_p\left(\norm{x_{k-N} - \bar{x}_{k-N}} \right) \right) + \\ &\quad\underline{\gamma}_p^{-1}\left(3 N \bar{\gamma}_w\left(\supofseq{w} \right) \right) + \underline{\gamma}_p^{-1}\left(3 N \bar{\gamma}_v\left(\supofseq{v} \right) \right), \\
&\leq \frac{3^{\frac{1}{a}}}{\norm{P_0^{-1}}} \left( \norm{P_{\infty}^{-1}}^{\frac{1}{a}} \norm{x_{k-N} - \bar{x}_{k-N}} + \right. \\
&\qquad\left. N^{\frac{1}{a}} \bar{\gamma}_w^{\frac{1}{a}}\left( \supofseq{w}\right) + N^{\frac{1}{a}} \bar{\gamma}_v^{\frac{1}{a}}\left( \supofseq{v}\right) \right).
\end{split}
\end{equation*}
Analogously, bounds for $\norm{\hat{v}_{j\vert k}}$ and $\norm{\hat{w}_{j\vert k}}$ can be found
\begin{equation}		\label{bound of w and v}
\begin{split}
\norm{\hat{w}_{j\vert k}} &\leq \underline{\gamma}_w^{-1}\left(\frac{3}{N}\bar{\gamma}_p\left(\norm{x_{k-N} - \bar{x}_{k-N}} \right) \right) + \\
&\qquad\underline{\gamma}_w^{-1}\left(3\; \bar{\gamma}_w\left(\supofseq{w} \right) \right) + \underline{\gamma}_w^{-1}\left(3\; \bar{\gamma}_v\left(\supofseq{v} \right) \right), \\
\norm{\hat{v}_{j\vert k}} &\leq \underline{\gamma}_v^{-1}\left(\frac{3}{N}\;\bar{\gamma}_p\left(\norm{x_{k-N} - \bar{x}_{k-N}} \right) \right) + \\
&\qquad\underline{\gamma}_v^{-1}\left(3\; \bar{\gamma}_w\left(\supofseq{w} \right) \right) + \underline{\gamma}_v^{-1}\left(3\; \bar{\gamma}_v\left(\supofseq{v} \right) \right).
\end{split}
\end{equation}
Next, let us consider some sample $k \in \mathbb{Z}_{\geq N}$. Assuming that system \eqref{eq_nonlinsys} is i-IOSS with $z_1 = x_{k-N}, z_2 = \hat{x}_{k-N \vert k}, w_1 = \{ w_{j} \}, w_2 = \{ \hat{w}_{j \vert k} \}, v_1 = \{ v_{j} \}$ and $v_2 = \{ \hat{v}_{j \vert k} \}$ for all $j \in \mathbb{Z}_{\left[k-N, k-1 \right]}$. Since $x(k)=x\left( N,z_1,\boldsymbol{w_1}  \right), \hat{x}(k)=\hat{x}_{k \vert k}=x\left( N,z_2,\boldsymbol{w_2}  \right)$ we obtain
\begin{equation}		\label{i-IOSS start development}
\begin{split}
  \norm{x_k - \hat{x}_{k\vert k}} &\leq 
\beta\left(\norm{x_{k-N} - \hat{x}_{k-N\vert k}}, \;N \right) + \gamma_1\left(\supofseq{w_j - \hat{w}_{j\vert k}} \right) \\
& \qquad  \quad + \gamma_2\left(\supofseq{v_j - \hat{v}_{j\vert k}} \right).
\end{split}
\end{equation}
In order to get a finite upper bound for the estimation error, the three terms in the right hand side of equation \eqref{i-IOSS start development} must be upper bounded. The first term can be written
\begin{equation*}
\begin{split}
\beta\left(\norm{x_{k-N} - \hat{x}_{k-N\vert k}}, \;N \right) \leq \beta\left(2\;\norm{x_{k-N} - \bar{x}_{k-N}}, \;N \right) + \\
\beta \left(2\;\norm{\hat{x}_{k-N\vert k} - \bar{x}_{k-N}}, \;N \right) \\
\leq \beta\left(2\norm{x_{k-N} - \bar{x}_{k-N}}, \;N \right) + \\
\beta \left( \frac{2\;3^{\frac{1}{a}} \norm{P_{\infty}^{-1}}^{\frac{1}{a}}  }{\norm{P_0^{-1}}} \norm{x_{k-N} - \bar{x}_{k-N}} + \frac{2\;3^{\frac{1}{a}} N^{\frac{1}{a}} }{\norm{P_0^{-1}}} \; \bar{\gamma}_w^{\frac{1}{a}}\left( \supofseq{w}\right) + \right.\\
\left. \frac{2\;3^{\frac{1}{a}} N^{\frac{1}{a}}}{\norm{P_0^{-1}}} \; \bar{\gamma}_v^{\frac{1}{a}}\left( \supofseq{v}\right), N \right) 	\\
\leq \beta\left(2 \; \norm{x_{k-N} - \bar{x}_{k-N}}, \;N \right) + \\ \beta\ \left( \frac{6\;3^{\frac{1}{a}} \norm{P_{\infty}^{-1}}^{\frac{1}{a}} }{\norm{P_0^{-1}}} \; \norm{x_{k-N} - \bar{x}_{k-N}},\;N \right) + \\
\beta\left(\frac{6\;3^{\frac{1}{a}} \;N^{\frac{1}{a}} }{\norm{P_0^{-1}}} \; \bar{\gamma}_w^{\frac{1}{a}}\left( \supofseq{w}\right),\;N\right) + \beta\left(\frac{6\;3^{\frac{1}{a}} \;N^{\frac{1}{a}} }{\norm{P_0^{-1}}}\; \bar{\gamma}_v^{\frac{1}{a}}\left( \supofseq{v}\right), N \right).
\end{split}
\end{equation*}
Using Assumptions \ref{prior weighting assumption} and \ref{assumption beta function ineq}, function $\beta (\cdot)$ is bounded by
\begin{equation*}
\begin{split}
\beta\left(\norm{x_{k-N} - \hat{x}_{k-N\vert k}} , N\right) \leq \frac{c_{\beta}\;2^p}{N^q} \norm{x_{k-N}-\bar{x}_{k-N}}^p +  \\
\frac{c_{\beta}\;6^p\;3^{\frac{p}{a}} \norm{P_{\infty}^{-1}}^{\frac{p}{a}}}{\norm{P_0^{-1}}^p\; N^q}  \; \norm{x_{k-N}-\bar{x}_{k-N}}^p +                              \\ \frac{c_{\beta}\;6^p\;3^{\frac{p}{a}}\;N^{\frac{p}{a}}}{\norm{P_0^{-1}}} \; \bar{\gamma}_w^{\frac{p}{a}}\left(\supofseq{w} \right) + \frac{c_{\beta}\;6^p\;3^{\frac{p}{a}}\;N^{\frac{p}{a}}}{\norm{P_0^{-1}}} \; \bar{\gamma}_v^{\frac{p}{a}}\left(\supofseq{v} \right)        \\
\leq \frac{c_{\beta}\;2^p}{N^q} \;\norm{x_{k-N}-\bar{x}_{k-N}}^p + \\
\left(\frac{\norm{P_{\infty}^{-1}}}{\norm{P_0^{-1}}}\right)^p \frac{c_{\beta}\;6^p\;3^{\frac{p}{a}}}{N^q}  \; \norm{x_{k-N}-\bar{x}_{k-N}}^p +\\
\frac{c_{\beta}\;6^p\;3^{\frac{p}{a}}\;N^{\frac{p}{a}-q} }{\norm{P_0^{-1}}} \; \bar{\gamma}_w^{\frac{p}{a}}\left(\supofseq{w} \right) + \frac{c_{\beta}\;6^p\;3^{\frac{p}{a}}\;N^{\frac{p}{a}-q} }{\norm{P_0^{-1}}} \; \bar{\gamma}_v^{\frac{p}{a}}\left(\supofseq{v} \right).
\end{split}
\end{equation*}

Taking in account that $P_k^{-1}$ is a symmetric positive definite matrix for all $k \in \mathbb{Z}_{\left[0, \infty \right)}$, then $\norm{P_k^{-1}} \leq \lambda_{\max}\left( P_k^{-1}\right)$, where $\lambda_{\max}\left(P_k^{-1} \right)$ denotes the maximal eigenvalue of matrix $P_k^{-1}$. Denoting $\lambda_{\min}\left(P_k^{-1} \right)$ as the minimal eigenvalue of matrix $P_k^{-1}$ and taking in account that $\norm{P_k^{-1}} \leq \norm{P_{k+1}^{-1}}$, the maximum conditioning number of matrix $P_k^{-1}$ can be defined as $\mathbb{C}_{P^{-1}} \coloneqq \lambda_{\max}\left(P_{\infty}^{-1} \right) / \lambda_{\min}\left(P_{0}^{-1} \right)$, then  $\beta\left(\norm{x_{k-N} - \hat{x}_{k-N\vert k}},\;N \right)$ can be bounded by
\begin{equation}		\label{bound of beta in i-IOSS prop.}
\begin{split}
\beta\left(\norm{x_{k-N} - \hat{x}_{k-N\vert k}} ,\; N\right) \leq 
\frac{c_{\beta}\;18^p}{\norm{P_0^{-1}}} \left( \bar{\gamma}_w^{\frac{p}{a}}\left(\supofseq{w} \right) + \bar{\gamma}_v^{\frac{p}{a}}\left(\supofseq{v} \right) \right)+ \\
\frac{c_{\beta}}{N^q} \left( 2^p + \mathbb{C}_{P^{-1}}^p 18^p \right)
\norm{x_{k-N}-\bar{x}_{k-N}}^p.
\end{split}
\end{equation}

The first term in the right side of this equation is bounded due the assumption that $\norm{x_{k-N} - \bar{x}_{k-N}} \in \mathcal{X}_0\left(e_{\max} \right)$, while the second term are finite constants. To extend the validness of \eqref{bound of beta in i-IOSS prop.} to the full estimation horizon, an extension of the function $\beta$ at the beginning of the estimation, $N = 0$, is required.

The second term in the right hand side of equation \eqref{i-IOSS start development}, can be bounded by the following inequality  $\forall j \in \mathbb{Z}_{\left[k-N, k-1 \right]}$

\begin{equation*}
\begin{split}
\gamma_1\left(\supofseq{w_j - \hat{w}_{j\vert k}} \right) &\leq  \gamma_1\left( \supofseq{w} + \supofseq{\hat{w}_{j\vert k}} \right)                            \\
&\leq \gamma_1\left(\supofseq{w} + \underline{\gamma}_w^{-1}\left(\frac{3}{N} \bar{\gamma}_p\left(\norm{x_{k-N} - \bar{x}_{k-N}} \right) \right) + \right.    \\
    & \qquad \left. \underline{\gamma}_w^{-1}\left( 3\; \bar{\gamma}_w\left(\supofseq{w} \right) \right) + \underline{\gamma}_w^{-1}\left(3\; \bar{\gamma}_v\left(\supofseq{v} \right) \right) \right).
\end{split}
\end{equation*}

Recalling Assumption \ref{stage cost assumption}, the reader can verify the following inequality
\begin{equation}		\label{bound of second term of i-IOSS}
\begin{split}
\gamma_1\left(\supofseq{w_j - \hat{w}_{j\vert k}} \right)   
    \leq \frac{c_1 3^{\alpha_1} \norm{P_{\infty}^{-1}}^{\alpha_1} }{N^{\alpha_1}} \; \norm{x_{k-N} - \bar{x}_{k-N}}^{a \alpha_1} + \\
    c_1 3^{\alpha_1} \bar{\gamma}_v^{\alpha_1}\left(\supofseq{v} \right)
	\quad + \gamma_1\left(3\left(\supofseq{w} + \underline{\gamma}_w^{-1}\left( 3 \bar{\gamma}_w\left(\supofseq{w} \right) \right) \right) \right).
\end{split}
\end{equation}

In an equivalent manner, a bound for the third term in the right hand side of equation \eqref{i-IOSS start development} can be found

\begin{equation}		\label{bound of third term of i-IOSS}
\begin{split}
\gamma_2\left(\supofseq{v_j - \hat{v}_{j\vert k}} \right) \leq \frac{c_2 \;3^{\alpha_2}\;\norm{P_{\infty}^{-1}}^{\alpha_2}}{N^{\alpha_2}} \norm{x_{k-N} - \bar{x}_{k-N}}^{a \alpha_2} + \\
c_2\; 3^{\alpha_2} \bar{\gamma}_w^{\alpha_2}\left(\supofseq{w} \right) + \gamma_2\left(3\left(\supofseq{v} + \underline{\gamma}_v^{-1}\left( 3 \bar{\gamma}_v\left(\supofseq{v} \right) \right) \right) \right).
\end{split}
\end{equation}

Once an upper bound for the three terms of equation \eqref{i-IOSS start development} were found, defining $\zeta \coloneqq \max\{p, a\alpha_1, a\alpha_2 \}, \eta \coloneqq \min\{q, \alpha_1, \alpha_2 \}$ and $\rho \coloneqq \max\{p, \alpha_1, \alpha_2\}$, equation \eqref{i-IOSS start development} can be rewritten as follows
\begin{equation}	\label{final bound of x_k-n - x_k-n_bar}
  \begin{split}
    \norm{x_k - \hat{x}_{k\vert k}} \leq \frac{\norm{x_{k-N} - \bar{x}_{k-N}}^{\zeta}}{N^\eta}\left(\mathbb{C}_{P^{-1}}^{\rho}\left(c_{\beta}\;18^p + \right.\right.\\
    \left.\left. c_1\;3^{\alpha_1}\;\lambda_{\min}^{\alpha_1}\left(P_0^{-1} \right) + c_2\;3^{\alpha_2}\;\lambda_{\min}^{\alpha_1}\left(P_0^{-1}  \right)\right) + c_{\beta}\;2^p \right) + \\
    \left( \frac{c_{\beta}\;18^p\;\bar{\gamma}_w^{\frac{p}{a}}\left(\supofseq{w} \right)}{\norm{P_0^{-1}}} + \gamma_1\left( 3\left( \supofseq{w} + \underline{\gamma}_w^{-1}\left( 3\bar{\gamma}_w\left(\supofseq{w} \right) \right) \right) \right) + \right.\\
    \left. c_2\;3^{\alpha_2}\bar{\gamma}_w^{\alpha_2}\left(\supofseq{w} \right) \right) + \left( \frac{c_{\beta}\;18^p\;\bar{\gamma}_v^{\frac{p}{a}}\left(\supofseq{v} \right)}{\norm{P_0^{-1}}} + \right. \\
    \left. \gamma_2\left( 3\left( \supofseq{v} + \underline{\gamma}_v^{-1}\left( 3\bar{\gamma}_v\left(\supofseq{v} \right) \right) \right) \right) + c_1\;3^{\alpha_1}\bar{\gamma}_v^{\alpha_1}\left(\supofseq{v} \right) \right).
\end{split}
\end{equation}

Defining the functions $\bar{\beta}\left(r,\;s\right), \phi_w\left(r \right)$ and $\phi_v\left(r \right)$ for all $r \geq 0$ and $s \; \in \mathbb{Z}_{\geq 1 }$ as follows

\begin{align}
\begin{split}
 \bar{\beta}\left(r,\;s\right) &\coloneqq \frac{r^{\zeta}}{s^\eta}\left(\mathbb{C}_{P^{-1}}^{\rho}\left(c_{\beta}\;18^p +\right.\right.\\
 & \quad \left.\left.\lambda_{\min}^{\alpha_1}\left(P_0^{-1}  \right) \left(c_1\;3^{\alpha_1} + c_2\;3^{\alpha_2}\;\right)\right) + c_{\beta}\;2^p \right),
 \label{beta bar}
\end{split}    \\
\begin{split}
 \phi_w\left(r \right) &\coloneqq \frac{c_{\beta}\;18^p\;\bar{\gamma}_w^{\frac{p}{a}}\left(\supofseq{r} \right)}{\norm{P_0^{-1}}} +
 \gamma_1\left( 3\left( \supofseq{r} + \underline{\gamma}_w^{-1}\left( 3\bar{\gamma}_w\left(\supofseq{r} \right) \right) \right) \right) + \\ & \qquad c_2\;3^{\alpha_2}\bar{\gamma}_w^{\alpha_2}\left(\supofseq{r}\right),
 \label{Phi_w}
\end{split}     \\
\begin{split}
 \phi_v\left(r \right) &\coloneqq \frac{c_{\beta}\;18^p\;\bar{\gamma}_v^{\frac{p}{a}}\left(\supofseq{r} \right)}{\norm{P_0^{-1}}} +
 \gamma_2\left( 3\left( \supofseq{r} + \underline{\gamma}_v^{-1}\left( 3\bar{\gamma}_v\left(\supofseq{r} \right) \right) \right) \right) + \\ & \qquad c_1\;3^{\alpha_1}\bar{\gamma}_v^{\alpha_1}\left(\supofseq{r} \right).
 \label{Phi_v}
\end{split}
\end{align}
 
equation \eqref{final bound of x_k-n - x_k-n_bar} can be written $\forall k \in \mathbb{Z}_{\left[1, N-1 \right]}$ as follows
 \begin{equation}		\label{final bound of x_k-n - x_k-n_bar_short}
  \norm{x_k - \hat{x}_{k\vert k}} \leq \bar{\beta}\left(\norm{x_{k-N} - \bar{x}_{k-N}},
  N \right) + \phi_w\left(\supofseq{w} \right) + \phi_v\left(\supofseq{v} \right) .
\end{equation}

To guarantee the validity of previous results on the entire time horizon we must extend the definition of $\beta\left( r, s\right) \text{ to } s = 0$. Because of $\bar{\beta}\left(r, s \right) \in \mathcal{KL}$, $\bar{\beta}\left(r, 0 \right) \in \mathcal{KL}$ and $\bar{\beta}\left(r, 0 \right) \geq \bar{\beta}\left(r, k \right)$ for $k \in \mathbb{Z}_{\geq 1}$, it is sufficient to define $\bar{\beta}\left(r, 0 \right) \geq k_{\beta} \; \bar{\beta}\left(r, 1 \right)$ for some $k_{\beta} \in \mathbb{R}_{> 1}$ to extend the definition of $\bar{\beta}\left(r, s \right) \text{for all } k \in \mathbb{Z}_{\geq 0}$.
We would like to determinate the decreasing rate for the function $\bar{\beta}\left(r, s \right)$ $\mathcal{N}$ samplings time in the future. In order to do that, let define the constants
\begin{equation*}
  \mu \in \mathbb{R}_{>0} \textnormal, \; \delta > \frac{2+\mu}{1+\mu}
\end{equation*}
and
\begin{eqnarray*}
    r_{\max} \coloneqq & \max \{\frac{1}{\delta}\left(\bar{\beta}
    \left(e_{\max}, \; 0 \right) + \phi_w\left( \supofseq{w}\right) + \phi_v
    \left( \supofseq{v}\right) \right),  \\ 
    & \qquad \quad \delta (1+\mu) \left(\phi_w \left(\supofseq{w} \right) +
    \phi_v\left(\supofseq{v} \right) \right) \} &
\end{eqnarray*}

The minimum horizon length required to accomplish a decreasing rate $\delta$ will be given by

\begin{equation}
\label{N minimum length}
\begin{array}{rl}
    \mathcal{N} \geq& \left(\delta^{\zeta} r_{\max}^{\zeta-1} \mathbb{C}_{P^{-1}}^{\rho}\left( c_{\beta} 18^p + \lambda_{\min}^{\alpha_1}\left( P_0^{-1} \right) \left( c_1 3^{\alpha_1} + \right.\right.\right. \\
    & \qquad \left.\left.\left.  c_2 3^{\alpha_2}\right) + c_{\beta} 2^p \right)\right)^{\frac{1}{\eta}}
\end{array}
\end{equation}

Adopting an estimator with a window length greater or equal to $\mathcal{N}$ such that
\begin{equation}		\label{beta equiv for N0}
  \bar{\beta}\left(\delta r, \; N \right) \leq \left(\frac{\mathcal{N}}{N}
  \right)^{\eta} r,
\end{equation}
the effects of the initial conditions will vanish with a decreasing rate $\delta$. As $k \rightarrow \infty$, the estimation will entry to the bounded set $\mathcal{X}\left(w,v \right) \in \mathcal{X}$ defined by the noises of the system

\begin{equation}
\begin{split}
  \mathcal{X} \left(w,v \right) &\coloneqq \{ \norm{x_{k+j} - 
  \hat{x}_{k+j\vert k+j}} \leq \delta \left(1 + \mu \right)\left( 
  \phi_w \left(\supofseq{w} \right) + \right.  \\
  &\qquad \left. \phi_v\left(\supofseq{v} \right) \right) \}.
\end{split}
\end{equation}

This set define the minimum size region of error space $\mathcal{X}$ that the error can achieve by removing the effect of errors in initial conditions ($e_{max}$). Equation \eqref{beta equiv for N0} establish a trade off between speed of convergence and window length, which is related with the size of $\mathcal{X}\left(w,v \right)$. 

For any MHE with adaptive arrival cost and window length $N \geq \mathcal{N}$ two situations can be considered
\begin{itemize}
  \item The estimator removed the effects of $x_0$ on $\hat{x}_{k+j\vert k+j}$ such that
  $x_{k+j}-\hat{x}_{k+j\vert k+j} \in \mathcal{X}\left(w,v \right)$, and
  \item The estimator has not removed the effects of $x_0$ on $\hat{x}_{k+j\vert k+j}$ such that $x_{k+j}-\hat{x}_{k+j\vert k+j}
  \notin \mathcal{X}\left(w,v \right)$,
\end{itemize}
Assuming the first situation and recalling equations \eqref{final bound of x_k-n - x_k-n_bar_short} and \eqref{beta equiv for N0}, the following inequalities hold
\begin{equation}		\label{first case invariant set}
  \begin{split}
    \norm{x_{k+N} - \hat{x}_{k+N\vert k+N}} &\leq \bar{\beta}\left(\norm{x_{k} - 
    \bar{x}_{k}}, \,k \right) + \phi_w\left(\supofseq{w} \right) + \\
    & \qquad \qquad \phi_v\left(\supofseq{v} \right), \\
	&\leq \frac{\norm{x_{k} - \bar{x}_{k}}}{\delta}\left(\frac{\mathcal{N}}
    {N}\right)^{\eta} + \phi_w\left(\supofseq{w} \right) + \\
    & \qquad \qquad \phi_v\left(\supofseq{v}\right), \\
    &\leq \left(2 + \mu \right)\left(\phi_w\left(\supofseq{w} \right) + \phi_v
    \left(\supofseq{v} \right) \right), \\
	&\leq \delta \left(1 + \mu \right)\left(\phi_w\left(\supofseq{w} \right) + \phi_v
    \left(\supofseq{v} \right) \right). 
  \end{split}
\end{equation}
This equation implies the fact that the estimation error $x_{k+j}-\hat{x}_{k+j\vert k+j} \in \mathcal{X}\left(w,v \right) \quad \forall j \in \mathbb{Z}_{\geq 0}$.

In the other case, when the estimation error is outside of $\mathcal{X}\left(w,v \right)$, equations \eqref{final bound of x_k-n - x_k-n_bar_short} and \eqref{beta equiv for N0} are recalled again and the following inequalities hold
\begin{equation}		\label{second case invariant set}
  \begin{split}
  \norm{x_{k+N} - \hat{x}_{k+N\vert k+N}} \leq & \frac{\norm{x_{k} - \bar{x}_{k}}}
  {\delta}\left(\frac{\mathcal{N}}{N}\right)^{\eta} + \phi_w\left(\supofseq{w} \right) + \\
  & \qquad \qquad \phi_v\left(\supofseq{v} \right), \\
  \leq& \frac{\norm{x_{k} - \bar{x}_{k}}}{\delta}\left(\frac{\mathcal{N}}
  {N}\right)^{\eta} + \frac{\norm{x_{k} - \bar{x}_{k}}}{\delta \left(1 + \mu \right)}
  \left(\frac{\mathcal{N}}{N}\right)^{\eta}, \\
  \leq& \norm{x_{k} - \bar{x}_{k}}\left(\frac{\mathcal{N}}{N}\right)^{\eta} 
  \left(\frac{2 + \mu}{\delta \left(1 + \mu \right)} \right).
  \end{split}
\end{equation}
Since $\delta > \frac{2 + \mu}{1 +\mu}$, then $\forall N \geq \mathcal{N}$ we have
\begin{equation}		\label{contractive_rate}
\theta \coloneqq \left(\frac{\mathcal{N}}{N}\right)^{\eta}\left(\frac{2 + \mu}{\delta \left(1 + \mu \right)} \right) < 1.
\end{equation}
Equations \eqref{second case invariant set} and \eqref{contractive_rate} reveal a contractive behaviour of the estimation error with $\theta$ as contraction factor. For some finite time $k^*$ the estimation error will decrease until $x_{k^*+j}-\hat{x}_{k^*+j\vert k^*+j} \in \mathcal{X}\left(w,v \right)$.

In an equivalent formulation, equations \eqref{first case invariant set} and \eqref{second case invariant set} put in evidence the existence of a positive invariant set and a Lyapunov like function for the proposed estimator. From equation \eqref{second case invariant set}, one can see that for the case that the estimation error belong to the set $\mathcal{X}\left(w, v \right)^C \cap \mathcal{X}$, the estimation error decreases in a factor of $\theta$ every $\mathcal{N}$ sampling time. Taking in account the general case in which $\norm{x_{k} - \hat{x}_{k\vert k}} \in \mathcal{X}$ for $k \in \mathbb{Z}_{\geq \mathcal{N}}$, following the same procedure as in \cite{muller2017nonlinear}, we could define $i \coloneqq \lfloor{\frac{k}{N}}\rfloor$ (where $\lfloor \cdot \rfloor$ denotes the floor function) and $j \coloneqq k\; \textit{mod} \;N$, therefore $k = iN + j$. Combining equations \eqref{first case invariant set} and \eqref{second case invariant set} and the fact that $\norm{x_j - \hat{x}_{j}} \leq \delta r_{\max}$ for $j \in \mathbb{Z}_{\left[ 0, N-1 \right]}$ one can obtain
\begin{equation}		\label{final bound of estimation error 1}
  \begin{split}
    \norm{x_{k} - \hat{x}_{k\vert k}} &\leq \max \{\norm{x_{j} - \bar{x}_{j}} \theta^i
    ,\; \delta \left(1 + \mu \right)\left(\phi_w\left(\supofseq{w} \right) +\right.\\
    & \left. \qquad\qquad \phi_v \left(\supofseq{v} \right) \right) \}, \\
    & \leq \max\{ \theta^i \left( \bar{\beta}\left( \norm{x_0 - \bar{x}_0}, j \right)
    + \phi_w\left( \supofseq{w} \right) + \right. \\
    & \left.\phi_v\left( \supofseq{v} \right) \right),\quad \delta\left( 1+\mu \right)\left( \phi_w\left( \supofseq{w} \right) + \phi_v
    \left( \supofseq{v} \right) \right) \}, \\
    & \leq \theta^i \bar{\beta}\left( \norm{x_0 - \bar{x}_0}, \; j \right) + \delta
    \left( 1+\mu \right)\left( \phi_w\left( \supofseq{w} \right) + \right. \\
    & \left. \qquad \qquad \phi_v\left(\supofseq{v} \right) \right), \\
    & \leq \Phi\left( \norm{x_0 - \bar{x}_0}, \; k \right)  + \\
    & \qquad \delta\left( 1+\mu \right) \left( \phi_w\left( \supofseq{w} \right) + \phi_v\left( \supofseq{v} \right) 
    \right).
  \end{split}
\end{equation}
where
\begin{equation*}
  \Phi\left( \norm{x_0 - \bar{x}_0}, \; k \right) \coloneqq \theta^i
  \bar{\beta}\left( \norm{x_0 - \bar{x}_0}, \; j \right)
  \quad j \in \left[0, N-1 \right].
\end{equation*}
Since $\bar{\beta}\left(r, s \right) \in \mathcal{KL}$, function $\Phi\left( \norm{x_0 - \bar{x}_0}, \; k \right)$ could increase in the steps $\mathbb{Z}_{\left[ iN-1, iN \right]}$ for $i \geq 1$ (recall definition in Equation \eqref{beta bar}). Therefore, define $\bar{\Phi}\left( \norm{x_0 - \bar{x}_0}, \; k \right)$ which is an upper bound for $\Phi\left( \norm{x_0 - \bar{x}_0}, \; k \right)$. Taking in account that noises at time $\geq k$ do not affect the estimation at time $k$, equation \eqref{final bound of estimation error 1} can be rewritten as

\begin{equation}		\label{final bound of estimation error 2}
\begin{split}
\norm{x_{k} - \hat{x}_{k\vert k}} \leq & \bar{\Phi}\left( \norm{x_0 - \bar{x}_0}, \; k \right)  + \delta\left( 1+\mu \right)\left( \phi_w\left( \supofseq{w} \right)_{\left[0, k-1 \right]} + \right. \\
& \qquad \left. \phi_v \left( \supofseq{v} \right)_{\left[0, k-1 \right]} \right).
\end{split}
\end{equation}
This equation is just equation \eqref{RGAS def property} with 
\begin{align}
   \Phi\left(\norm{x_0 - \bar{x}_0 }, k \right) &= \bar{\Phi}\left( 
   \norm{x_0 - \bar{x}_0 }, \; k\right), \\
   \pi_w\left(\supofseq{w}_{\left[0, k-1 \right]} \right) &= \delta
   \left(1+\mu \right)\phi_w\left(\supofseq{w}_{\left[0, k-1 \right]}
   \right),			\label{Pi_w} \\
   \pi_v\left(\supofseq{v}_{\left[0, k-1 \right]} \right) &= \delta
   \left(1+\mu \right)\phi_v\left(\supofseq{v}_{\left[0, k-1 \right]}
   \right),			\label{Pi_v} 
\end{align}
therefore the estimator proposed in equations \eqref{eq:mhe1} is RGAS.

Finally, in order to prove that the estimation error $\norm{x_k - \hat{x}_{k\vert k}} \rightarrow 0$ when $\norm{w_{k}} \in \mathcal{C}_w, \norm{v_{k}} \in \mathcal{C}_v$, we must note that equation \eqref{final bound of x_k-n - x_k-n_bar_short} holds for $\supofseq{w}_{\left[k-N, k-1 \right]}$ instead of $\supofseq{w}$ and $\supofseq{v}_{\left[k-N, k-1 \right]}$ instead $\supofseq{v}$ (it can be done omitting last step in equation \eqref{upper bound for cost function}). From a qualitative point of view, taking in account that function $\bar{\Phi}\left( \norm{x_0 - \bar{x}_0}, \; k \right) \in \mathcal{KL}$ and sequences $\boldsymbol{w}$ and $\boldsymbol{v}$ are convergent, the right hand side of equation \eqref{final bound of estimation error 2} tends to zero as $k \rightarrow \infty$. 
{\hfill$\square$}

The proof of Theorem \ref{Theorem_1} is constructive and provides an estimate of the estimation horizon $N$ required to guarantee RGAS of the MHE proposed in this work. The estimates $\mathcal{N}$ and functions $\Phi$, $\pi_{w}$ and $\pi_{v}$ can be quite conservative, since their derivation involved conservative estimates of noises, errors, stage costs and arrival cost. 

Note that the minimum horizon necessary to guarantee RGAS $\mathcal{N}$ depends on $r_{max}$, which depends on the class of disturbances considered (upper bounds of noises and error), the initial value of the prior weighting matrix $P_{0}$ and the bounds of the stage cost. The minimum horizon length is independent of $\supofseq{w}$, $\supofseq{v}$, and the same $\mathcal{N}$ ensures RGAS for all bounded disturbances and bounded prior error, like the result obtained by \cite{muller2017nonlinear}). This implies that we can prove the RGAS property for full information estimator with least--square objective function.
\begin{remark}
Functions $\phi_w$ and $\phi_v$ in equations \eqref{Phi_w} and \eqref{Phi_v}, and hence $\pi_w$ and $\pi_v$ in equations \eqref{Pi_w} and \eqref{Pi_v}, do not depend on the estimation horizon $N$ which means that the moving horizon estimator with adaptive arrival cost is RGAS with uniform gains given by \eqref{Phi_w} and \eqref{Phi_v}.
\end{remark}


\section{Examples}
\label{section:Examples}
The following examples will be used to illustrate the results presented in the previous sections and compare the performance of the estimators. The examples considered in this work are taken from \cite{muller2017nonlinear} for a direct comparison of the results.

\subsection{Example 1}
The first example considers the system
\begin{align}
    x(t+1) &= \left[ \begin{array}{ll}
         0.8 x_0(t) + 0.2 x_1(t) + 0.5 w(t)  \\
         -0.3 x_0(t) + 0.5 \cos(x_1(t))
    \end{array} \right] \\
    y(t) &= x_1(t) + v(t) \nonumber
\end{align}
The stage cost is chosen as $\ell(w,v) = 10 w^2 + 10 v^2$ and the horizon length is $N=10$. The prior weighting is chosen as $\Gamma(\chi) = 0.1 (\chi - \hat{x}(t|t))^T (\chi - \hat{x}(t|t))$ for the MAX estimator (\cite{muller2017nonlinear}) and $\Gamma_t(\chi) = (\chi - \hat{x}(t|t))^T \Pi_k^{-1} (\chi - \hat{x}(t|t))$ for the ADAP estimator (our method), where $\Pi_0=10I_2$ and $\Pi_k$ is obtained using equations \eqref{eq:updatePk} with $\sigma = 0.2$ and $c=1e6$. The MAX estimator uses $\delta = 1$, $\delta_1 = \kappa^N$ with $\kappa = 0.89^2$ and $\delta_2 = 1/N$ (see equation (3) of \cite{muller2017nonlinear}). The full information estimator (FIE MAX, see \cite{ji2016robust}) is configured with the same parameter used by \cite{muller2017nonlinear}, maintaining the stage cost and prior weighting $\Gamma_0$, and $\delta = 1$, $\delta_1 = \kappa^t$ and $\delta_2 = 1/t$.

\begin{table}[thb!]	
    \caption{Example 1 averaged MSE over 300 trials.}
    \vspace{1mm}
    \begin{tabularx}{\linewidth}{l|cccc}                                                \hline
                & \quad FIE MAX & \qquad ADAP & \qquad MAX & \qquad EKF            \\  \hline
        $x_0$ & \quad 0.02040 & \qquad 0.02176 & \qquad 0.02206 & \qquad 0.02296   \\ 
        $x_1$ & \quad 0.00135 & \qquad 0.00151 & \qquad 0.00156 & \qquad 0.00154   \\  \hline
    \end{tabularx}
\label{tab:example_1_avg_mse}
\end{table}

Table \ref{tab:example_1_avg_mse} shows the mean square estimation error of each estimator averaged over 300 trials. It can be seen that the proposed estimator average mean square estimation error is smaller than MAX ones and closer to FIE MAX. The main performance difference between ADAP and FIEMAX estimators is the inclusion of the \textit{max} term in the last one, which allows to follow the sudden changes (see Figures \ref{fig:ex1_x0} and \ref{fig:ex1_x1}).   

\begin{figure}[!]
  \begin{tabular}{c}
        {
        \includegraphics[width=0.45\textwidth]{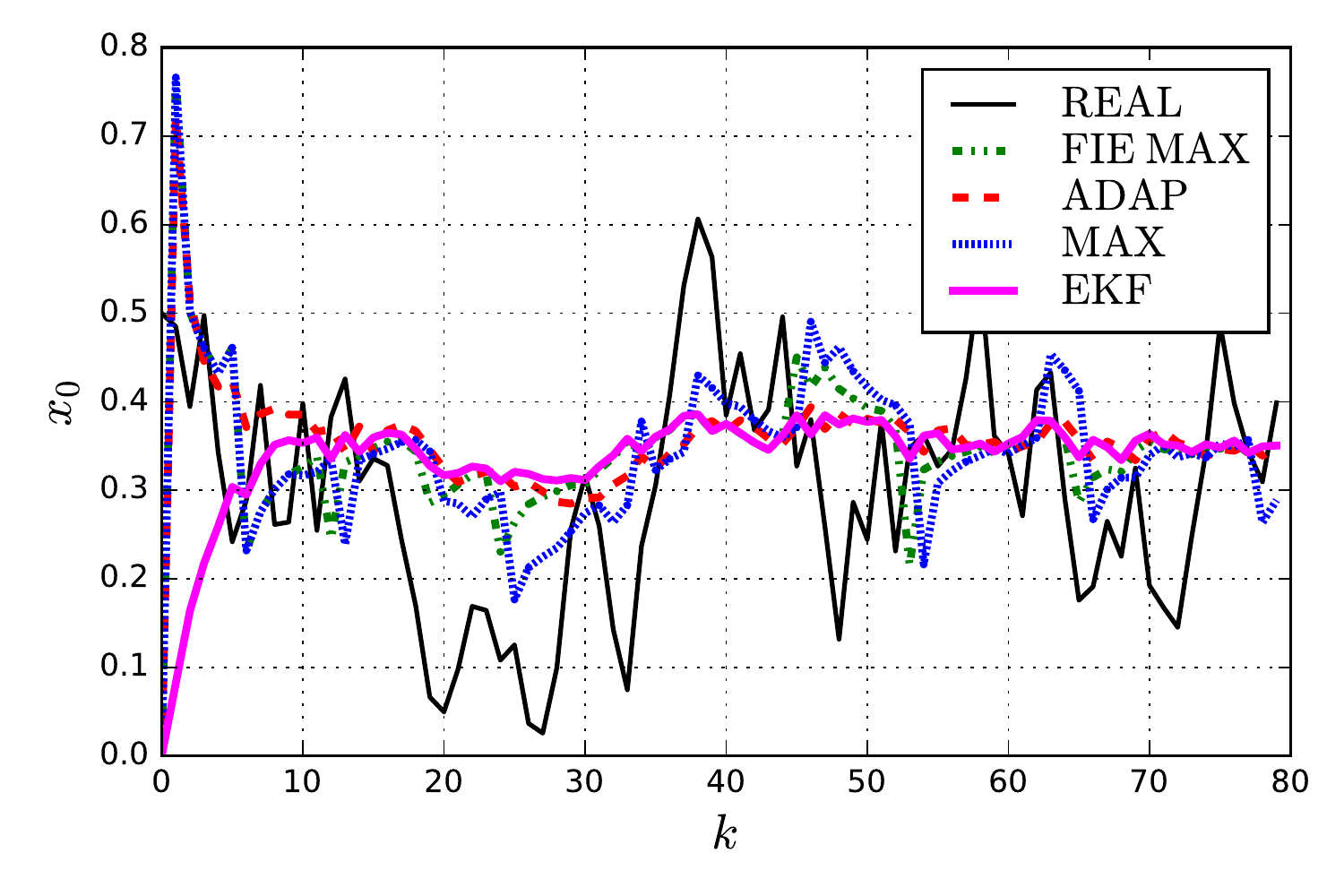}
        \label{fig:ex1_x0}}  \\
  \end{tabular}
  \caption{Comparison between ADAP (red dash dotted), MAX (blue dashed), FIEMAX (green dotted), EKF (magenta) estimators, and real system state (black solid).}
  \label{fig:ex1_x0}
\end{figure}

Figures \ref{fig:ex1_x0} and \ref{fig:ex1_x1} shows simulation results with initial condition $x_0 = [0.5, 0]^T$ and prior estimate $\bar{x}_0 = [0, 0]^T$. The process and measurement disturbances $w$ and $v$ are sampled from an uniform distribution over the intervals $[-0.3, 0.3]$ and $[-0.2, 0.2]$, respectively. This figure shows that the estimators that use the \textit{max} term are able of following the sudden changes, however in the remaining of the signal the MAX estimator is moving away of the FIEMAX while ADAP remains closer.

\begin{figure}[!]
  \begin{tabular}{c}
        {
        \includegraphics[width=0.45\textwidth]{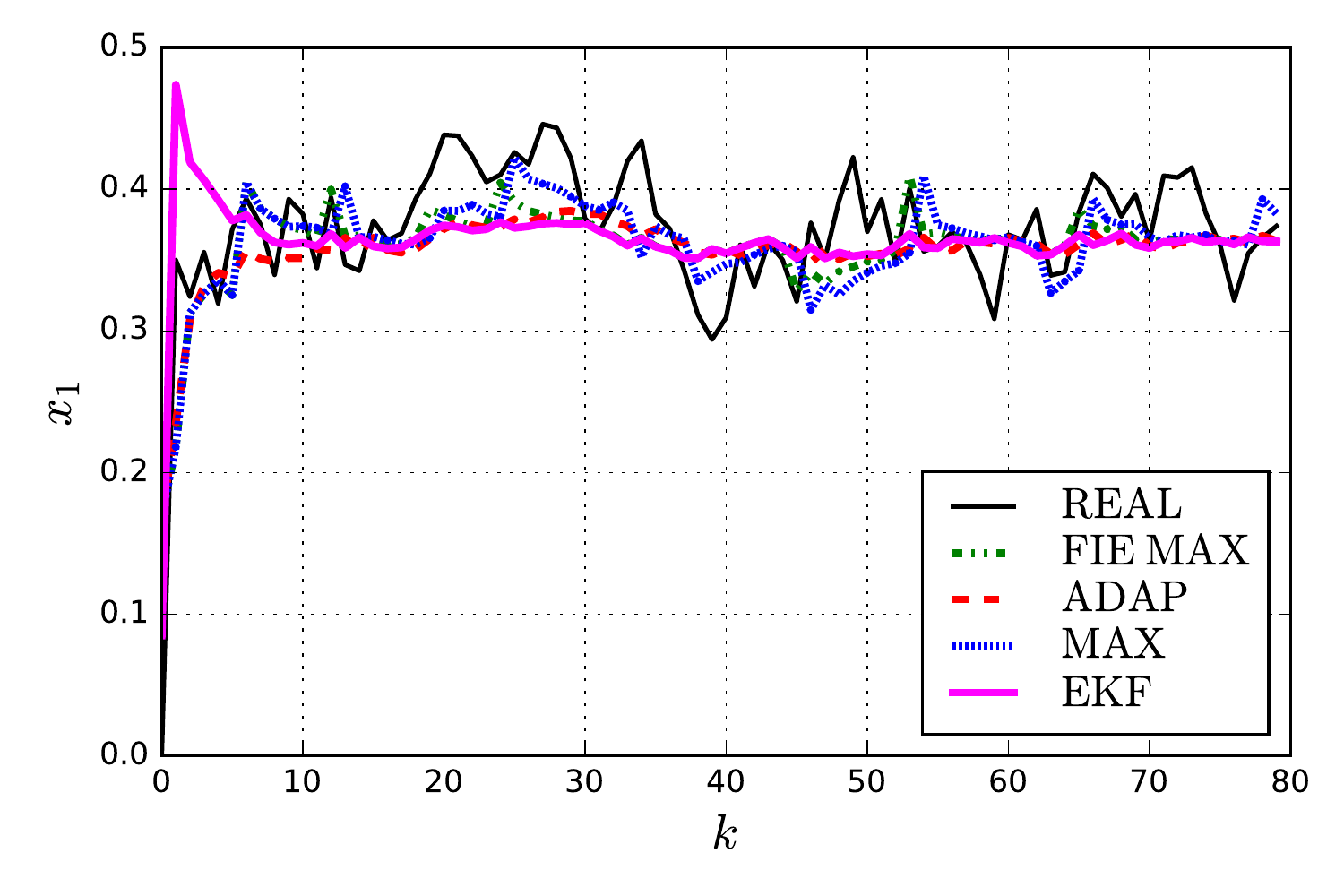}
        \label{fig:ex1_x1}}
  \end{tabular}
  \caption{Comparison between ADAP (red dash dotted), MAX (blue dashed), FIEMAX (green dotted), EKF (magenta) estimators, and real system state (black solid).}
  \label{fig:ex1_x1}
\end{figure}

\subsubsection{MHE in the presence of variable measurement noise}
Now the MHE estimator is evaluated in the presence of time-varying measurement noise. The variance of the measurement noise is changed from $0.2$ to $1.0$ between times 20 and 40, then it returns to $0.2$.

\begin{table}[thb!]
    \caption{Example 1 aver. MSE over 300 trials with variable measurement noise.}
    \vspace{1mm}
	\begin{tabularx}{\linewidth}{l|ccc}                                     \hline
   		& \quad\qquad  ADAP & \qquad MAX & \qquad FIE MAX                      \\\hline
		$x_0$  & \quad\qquad 0.02068 & \qquad  0.03067 & \qquad  0.00761       \\ 
		$x_1$  & \quad\qquad 0.00290 & \qquad  0.00335 & \qquad  0.00068       \\\hline
	\end{tabularx}
	\label{tab:example_1_avg_mse_vn}	
\end{table}
Table \ref{tab:example_1_avg_mse_vn} shows the average mean square error in the presence of variable measurement noise. In this case we can see that the behaviour of the proposed estimator is marginally affected by the variations of the measurement, while the mean square error of $x_0$ of other estimators increase significantly. These behaviours are due to the adaptation capabilities of the prior weighting updating mechanism, which is able of tracking the changes of noises, in the case of ADAP estimator, and the effect of the \textit{max} term in MAX and FIEMAX estimators.

\begin{figure}[bp!]
    \centering
    \includegraphics[width=0.45\textwidth]{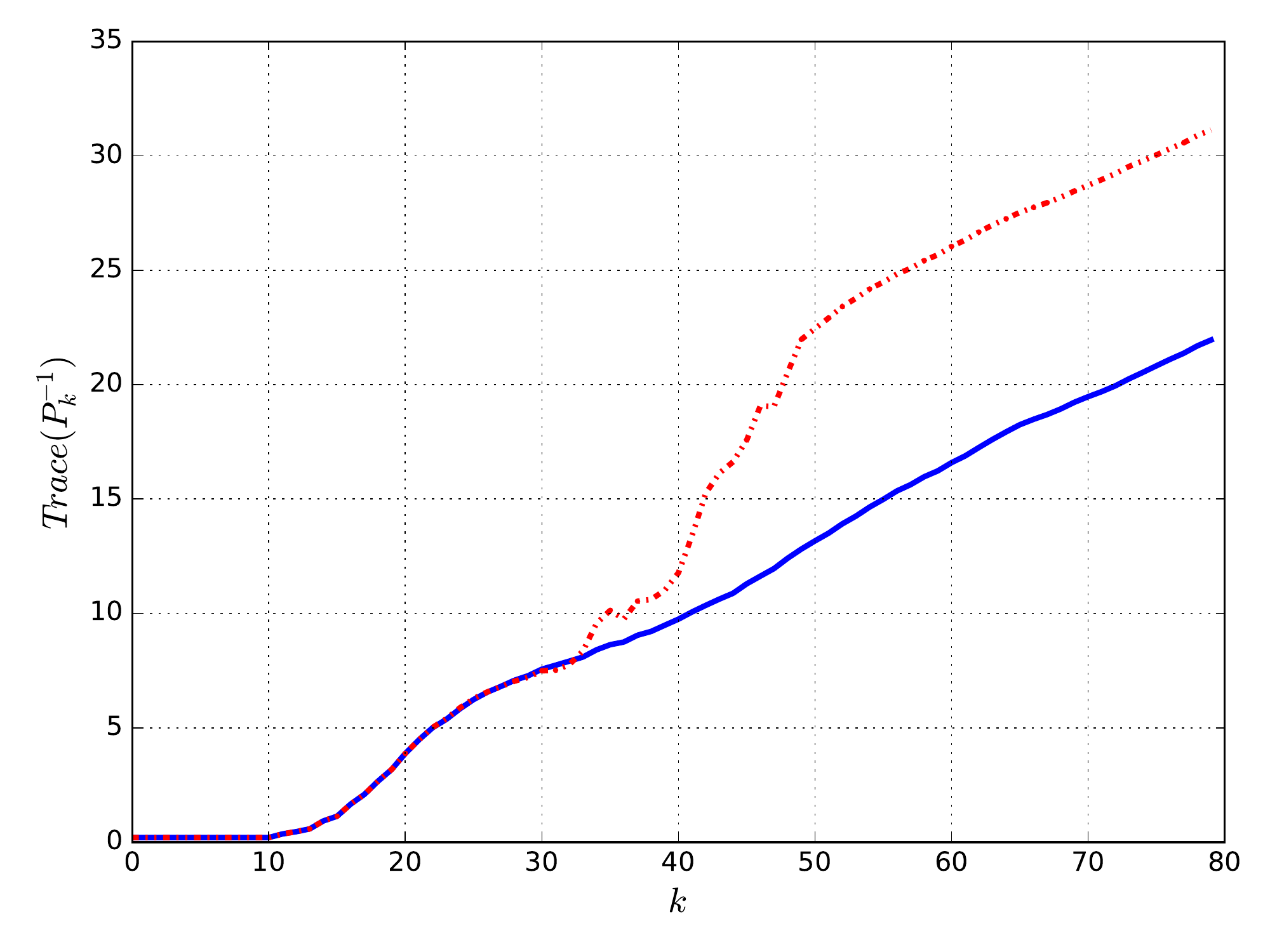}
    \caption{Comparison of the evolution of $trace(P^{-1}_{k-N})$ used by ADAP estimator for time--varying (red dash dotted) and constant (blue dashed) measurement noise parameters.}
    \label{fig:TrInvP_vn}
\end{figure}
Figure \ref{fig:TrInvP_vn} shows the evolution of the trace of $P^{-1}_{k-N}$ used in the prior weight of ADAP estimator in both examples. It can be seen that the trace of both matrices grow in similar way, however when the measurement noise changes its variance from $0.2$ to $1.0$ the trace of $P^{-1}_{k-N}$ increases its value (from $12.5$ to $22.5$) and them both traces have the same behaviour.

\subsection{Example 2}
As a second example, we consider a second order gas-phase irreversible reaction of the form $2A \rightarrow B$. This example has been considered in the context of moving horizon estimation in \cite{haseltine2005critical},  \cite{ji2016robust} and \cite{muller2017nonlinear}. Assuming an isothermal reaction and that the ideal gas law holds, the system dynamics
\begin{equation}
  \begin{split}
  \dot{x} &= \left[ \begin{array}{rc}
                   -2 k x_0^2  \\
         k x_0^2
    \end{array} \right] \\
    h(x) &= x_0 + x_1
  \end{split}
\end{equation}
where $x = [x_0, x_1]$, $x_0$ is the partial pressure of the reactant $A$, $x_1$ is the partial pressure of the product $B$, and $k=0.16$ is the reaction rate constant. The measured output of the system is the total pressure. The system is affected with additive process and measurement noise $w$ and $v$ drawn from normal distributions with zero mean and covariance $Q_w = 0.001^2 I_2$ and $R_v = 0.1^2$, respectively. The stage cost and prior weighting are chosen as $\ell(w,v) = w^T Q_w^{-1} w + R_v^{-1} v^2$ and $\Gamma_t(\chi) = (\chi - \hat{x}(t|t))^T \Pi_k^{-1} (\chi - \hat{x}(t|t))$ with $\Pi_0 = (1/36) I_2$, where $\Pi_k$ is determined by an extended Kalman filtering recursion in the case of the MAX estimator and the adaptive method in the case of the ADAP estimator with $\sigma = 0.1$ and $c=1e6$. For the MAX estimator we use $\delta_1=1/N$, $\delta_2=1$ and $\delta=0$. In the case of the ADAP estimator, the stage cost weight matrices are chosen as $Q_w = 0.001 I_2$ and $R_v = 0.1$. We use a multiple shooting strategy with a sampling time of $\Delta = 0.1$ and we add the restrictions $x_0 \geq 0$ and $x_1 \geq 0$. 

\begin{table}[thb]
    \centering
    \caption{Example 2 averaged MSE over 300 trials and different horizon size.}
    \vspace{1mm}
    \resizebox{0.47\textwidth}{!}{%
    \setlength{\extrarowheight}{3pt}
    \begin{tabular}{l|cc|cc|cc|c} \hline
    & \multicolumn{2}{c}{N=2} & \multicolumn{2}{|c}{N=5} & \multicolumn{2}{|c|}{N=10} & \\
      & ADAP & MAX & ADAP & MAX & ADAP & MAX & FIE \\ \hline
$x_0$ & 0.18808 & 0.58652 & 0.03367 & 0.04615 & 0.00171 & 0.00772 & 0.00024 \\ 
$x_1$ & 0.23037 & 0.66768 & 0.04074 & 0.05077 & 0.00285 & 0.00951 & 0.00120 \\ \hline
	\end{tabular}}
    \label{tab:example_2_avg_mse}
\end{table}
Table \ref{tab:example_2_avg_mse} shows the values of the mean squared error computed from the time $10$ (in order to neglect the initial transient error) up to the simulation end time and averaged over 300 trials for horizon sizes of $N=5$ and $N=10$.

\begin{figure}[!]
  \centering
  \begin{tabular}{c}
    {
    \includegraphics[width=0.45\textwidth]{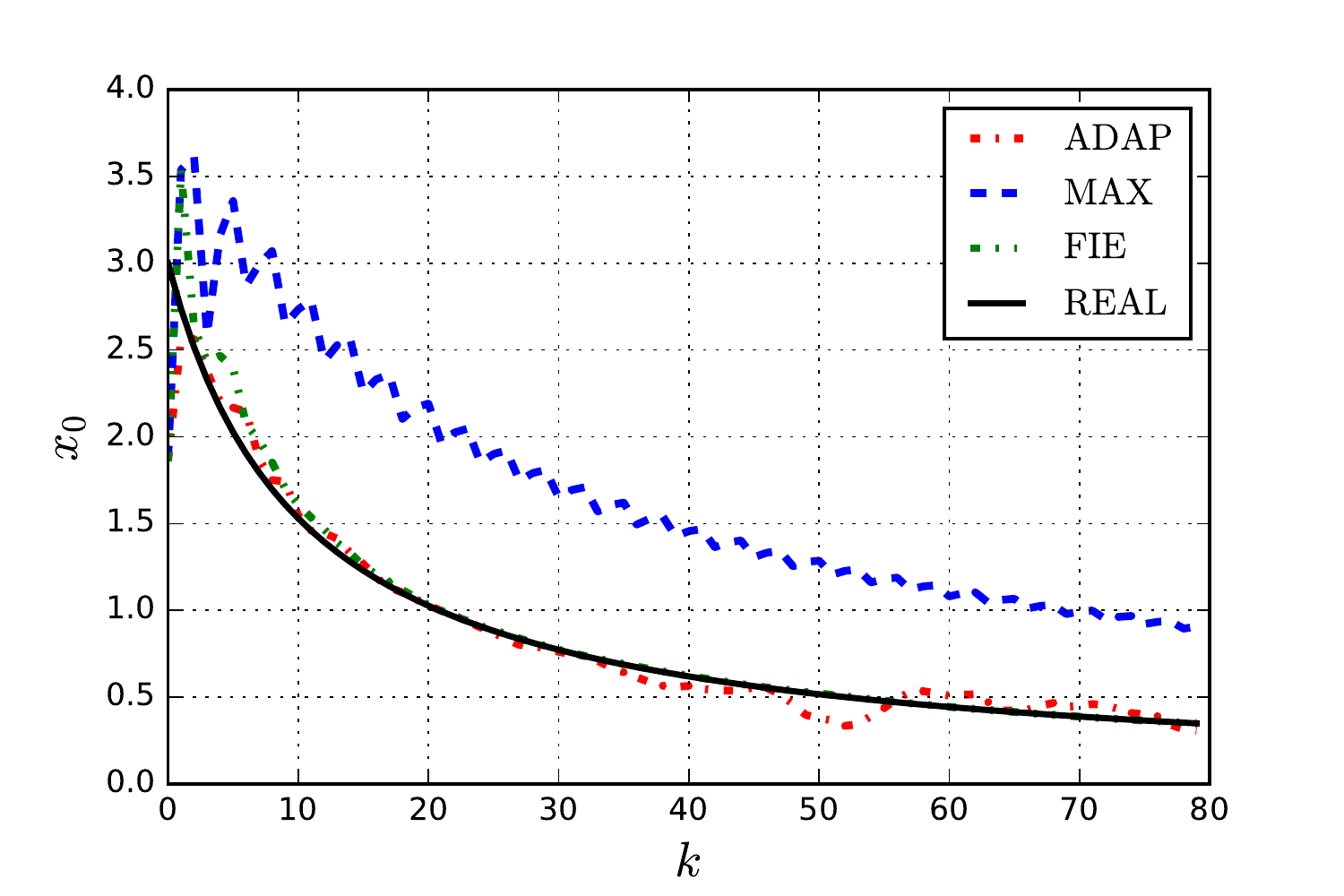}} \\
    {
    \includegraphics[width=0.45\textwidth]{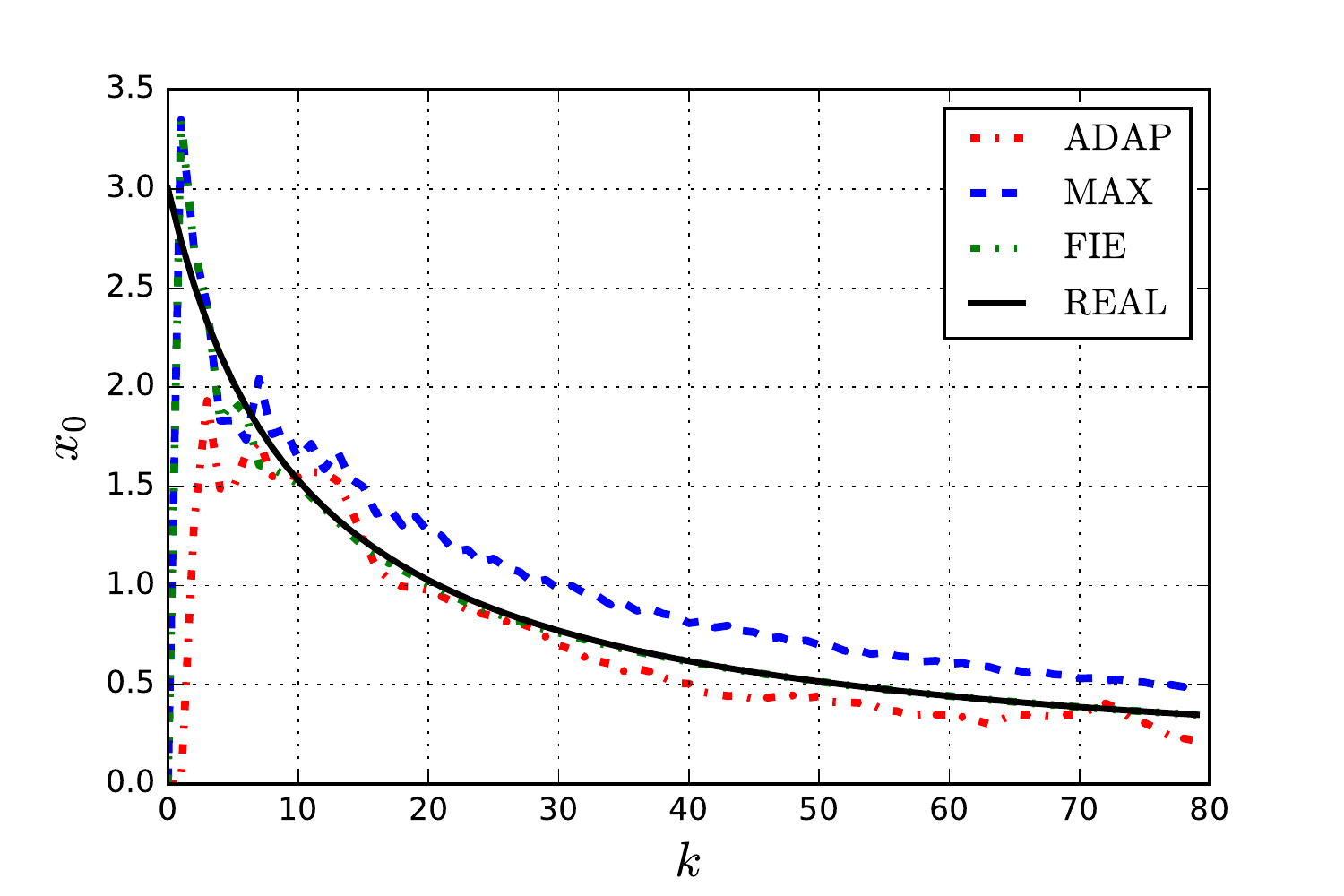}} \\
    {
    \includegraphics[width=0.45\textwidth]{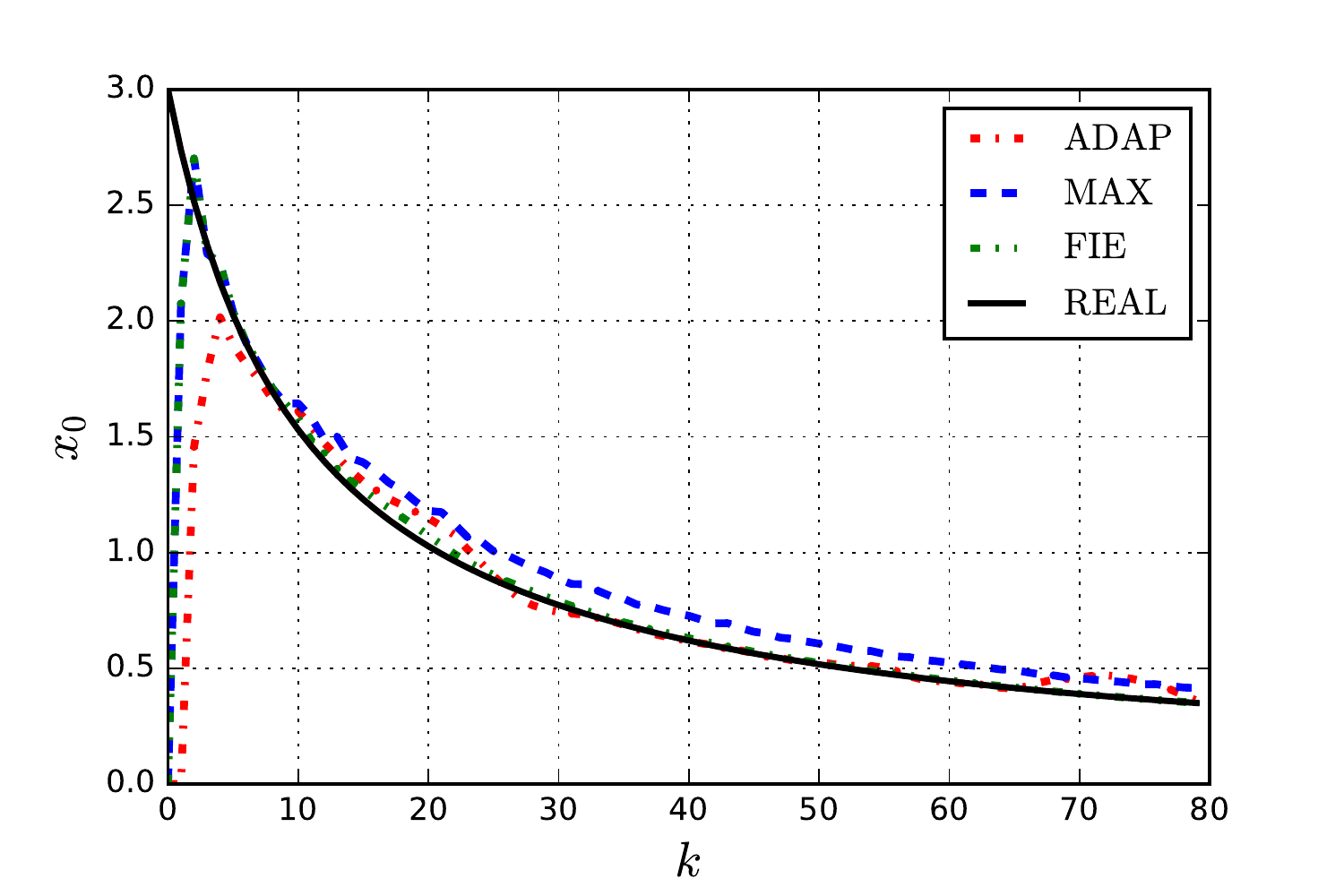}} \\
  \end{tabular}
  \caption{Comparison between ADAP (red dash dotted), MAX (blue dashed), FIEMAX (green dotted) estimators and real system state (black solid) for different horizon length ($N=2,5 \text{ and } 10$).}
 \label{fig:ex2_x0_nt}
\end{figure}

\begin{figure}[!]
  \centering
  \begin{tabular}{c}
    {
    \includegraphics[width=0.45\textwidth]{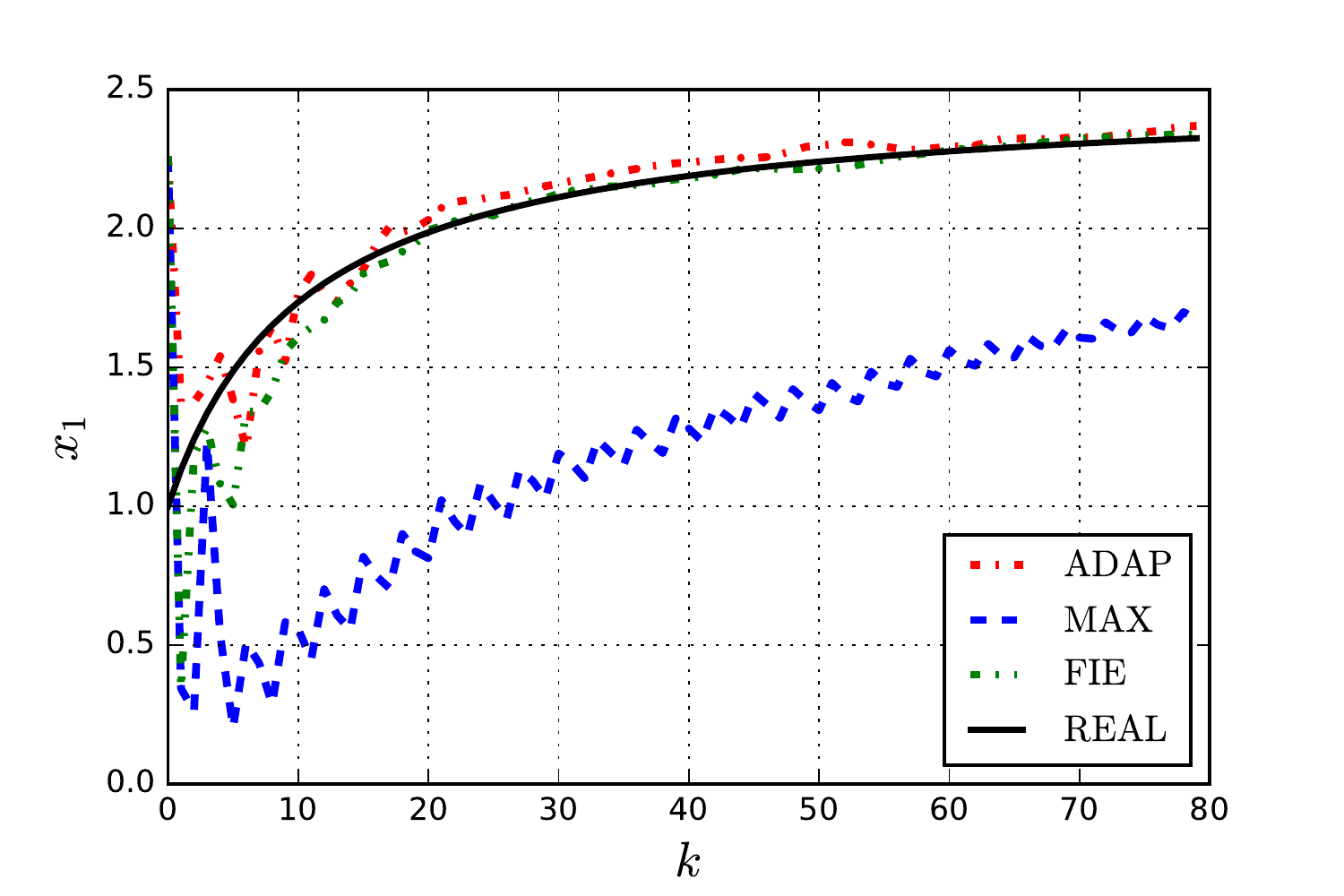}} \\
    {
    \includegraphics[width=0.45\textwidth]{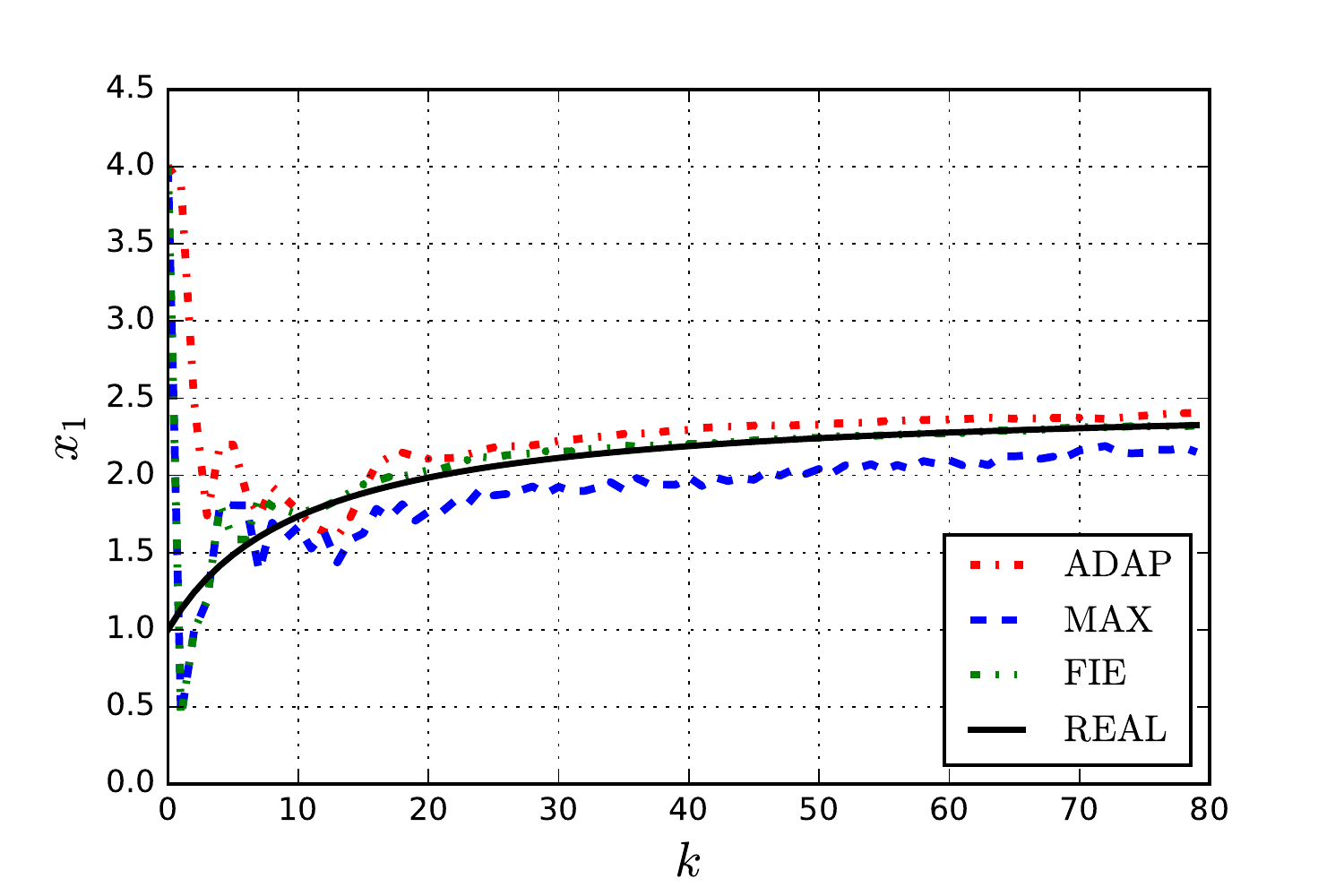}} \\
    {
    \includegraphics[width=0.45\textwidth]{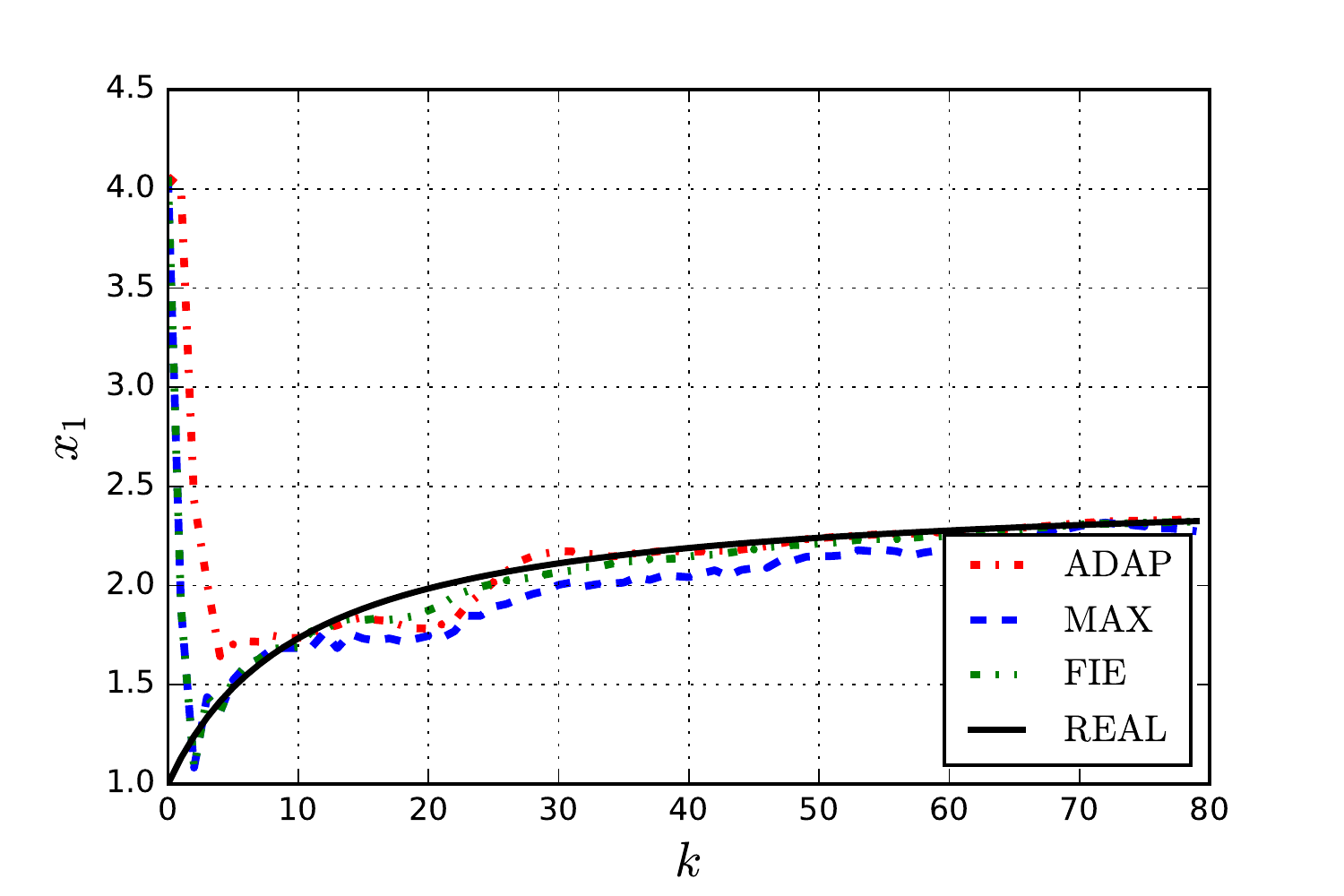}} \\
  \end{tabular}
  \caption{Comparison between ADAP (red dash dotted), MAX (blue dashed), FIEMAX (green dotted) estimators and real system state (black solid) for different horizon length ($N=2,5 \text{ and } 10$).}
 \label{fig:ex2_x1_nt}
\end{figure}

Figures \ref{fig:ex2_x0_nt} and \ref{fig:ex2_x1_nt} show simulation results with $x_0 = [3, 1]^T$ and $\bar{x}_0 = [0.1, 4.5]^T$ and horizons of sizes $N=2, 5 \text{ and } 10$, along with results for a full information estimator using the same parameters: the same stage cost $\ell(\cdot)$, prior weighting $\Gamma_0$, $\delta = 0$, $\delta_1 = 1/t$ and $\delta_2 = 1$. These figures show that the behaviour of ADAP estimator hardly change with horizon length (only the startup behaviours show differences) and no offset in the estimates, while the behaviour of the MAX estimator changes significantly. In addition to the cycling effect caused by the use of the filtered estimate to update $\bar{x}_{k-N}$ \citep{findeisen1997moving}, the MAX estimator also exhibits offset in the estimate that depends on the estimation horizon length.

\section{Conclusions}
In this paper we established robust global asymptotic stability for moving horizon estimator with a least-square type cost function for nonlinear detectable (i-IOSS) systems in presence of bounded disturbances. It was also shown that the estimation error converges to zero in case that disturbances converge to zero. This was done for an estimator which uses a least-square type cost function whose arrival cost us updated using adaptive estimation methods. An advantage of this updating mechanism is that the required conditions on prior weighting are such that it can be chosen off-line. Furthermore, it introduces a feedback mechanism between the arrival cost weight and the estimation errors that automatically controls the amount of information used to compute it, which allows to shorten the estimation horizon.

The standard least-square type cost function is typically used in practical applications and RGAS has been proved in \cite{muller2017nonlinear}. However, for this formulation, the disturbances gains depend on the estimation horizon. Hence, this result does not allow to establish robust global asymptotic stability for a full information estimator. We showed that changing the updating mechanism of arrival cost weight the disturbances gains becomes uniform, allowing to extend the stability analysis to full information estimators with least-square type cost functions.

\section*{Acknowledgment}
The authors wish to thank the Consejo Nacional
de Investigaciones Cientificas y Tecnicas (CONICET)
from Argentina, for their support. 


%





\bibliographystyle{unsrt}
\clearpage
\bibliography{sample.bib}

\end{document}